 \documentclass[final,5p,times,twocolumn]{elsarticle}

\usepackage{amssymb}
\usepackage{lipsum}
\usepackage{amsmath}

\usepackage{hyperref}
\usepackage{url}
\usepackage{ulem}
\usepackage{xcolor}
\biboptions{numbers,sort&compress}

\newcommand{\cu}{$^{76}$Cu}

\newcommand{\nuc}[2]{$^{#2}$#1}
\newcommand{\tran}[2]{$#1^+ \rightarrow #2^+$}

\journal{Physics Letters B}

\begin{document}

\begin{frontmatter}



\title{The \cu{} conundrum remains unsolved}


\author[iem,triumf,cern]{B. Olaizola}
\ead{bruno.olaizola@csic.es}
\author[ucm,jyu,hip,lnl]{A.~Illana}\ead{andres.illana@ucm.es}
\author[ucm]{J.~Benito}\ead{jabenito@ucm.es}
\author[IGFAE]{D.~P.~Su\'arez-Bustamante}
\author[cern] {G.~Del~Piccolo}

\author[ific]{A.~Algora}
\author[brat]{B.~Andel}
\author[york]{A.~N.~Andreyev}
\author[wars]{M.~Araszkiewicz}
\author[IGFAE]{Y.~Ayyad}
\author[ithemba]{R.~A.~Bark}
\author[surr]{T.~Berry}
\author[cern,iem]{M.~J.~G.~Borge}
\author[cern]{K.~Chrysalidis}
\author[kul]{T.~E.~Cocolios}
\author[nipne]{C.~Costache}
\author[cern,york]{J.~G.~Cubiss}
\author[kul]{P.~Van~Duppen}
\author[cern]{Z.~Favier}
\author[ucm]{L.~M.~Fraile}
\author[aarh]{H.~O.~U.~Fynbo}
\author[lnl]{F.~Galtarossa}
\author[ijc]{G.~Georgiev}
\author[jyu]{P.~T.~Greenless}
\author[tenn]{R.~Grzywacz}
\author[lib]{L.~J.~Harkness-Brennan}
\author[cern]{R.~Heinke}
\author[kul]{M.~Huyse}
\author[ucm]{P.~Iba\~nez}
\author[cern]{K.~Johnston}
\author[ithemba]{P.~M.~Jones}
\author[lib]{D.~S.~Judson}
\author[cern,jyu]{J.~Konki}
\author[wars]{A.~Korgul}
\author[ill]{U.~K\"oster}
\author[cern]{J.~Kurcewicz}
\author[stfc]{M.~Labiche}
\author[stfc]{I.~Lazarus}
\author[cern,nipne]{R.~Lic\u{a}}
\author[ucm]{M.~Llanos-Exp\'osito}
\author[cern,tenn]{M.~Madurga}
\author[nipne]{N.~Marginean}
\author[nipne]{R.~Marginean}
\author[cern]{B.~A.~Marsh\fnref{dead}}
\author[nipne]{C.~Mihai}
\author[nipne,czech]{R.~E.~Mihai}
\author[ucm]{J.~R.~Murias}
\author[iem,ific]{E.~N\'{a}cher}
\author[nipne]{C.~Neacsu}
\author[nipne]{A.~Negret}
\author[ucm]{V.~M.~Nouvilas}
\author[jyu]{J.~Ojala}
\author[west]{J.~N.~Orce}
\author[cern,york]{C.~A.~A.~Page}
\author[lib]{R.~D.~Page}
\author[jyu,hip]{J.~Pakarinen}
\author[stfc]{J.~Papadakis}
\author[nipne]{S.~Pascu}
\author[iem]{A.~Perea}
\author[wars,cern]{M.~Piersa-Siłkowska}
\author[lib,jyu,hip]{A.~M.~Plaza}
\author[surr,cern]{Zs.~Podoly\'ak}
\author[wars]{W.~Poklepa}
\author[stfc]{V.~Pucknell}
\author[jyu]{P.~Rahkila}
\author[york]{C.~Raison}
\author[cern]{E.~Rapisarda}
\author[kul]{K.~Rezynkina}
\author[nipne]{F.~Rotaru}
\author[koln]{K.~Schomacker}
\author[lnl,argo]{M.~Siciliano}
\author[nipne]{C.~Sotty}
\author[kul,jyu,hip]{M.~Stryjczyk}
\author[iem]{O.~Tengblad}
\author[ucm]{J.~M.~Ud\'{i}as}
\author[ucm]{V.~Vedia}
\author[iem]{S.~Vi\~nals}
\author[york]{R.~Wadsworth}
\author[koln]{N.~Warr}
\author[kul]{H.~De~Witte}
\author[triumf,ubc]{D.~Yates}
\author[cern,york]{Z.~Yue}
\author{and the IDS collaboration}

\address[iem]{Instituto de Estructura de la Materia, CSIC, 28006 Madrid, Spain}
\address[triumf]{TRIUMF, 4004 Wesbrook Mall, Vancouver, British Columbia V6T 2A3, Canada}
\address[cern]{CERN, CH-1211 Geneva, Switzerland}
\address[ucm]{Grupo de F\'isica Nuclear, EMFTEL \& IPARCOS, Universidad Complutense de Madrid, CEI Moncloa, E-28040 Madrid, Spain}
\address[jyu]{Accelerator Laboratory, Department of Physics, University of Jyv\"{a}skyl\"{a}, P.O. Box 35, FI-40014 University of Jyv\"askyl\"a, Finland}
\address[hip]{Helsinki Institute of Physics, University of Helsinki, P.O. Box 64, FI-00014 Helsinki, Finland}
\address[lnl]{Istituto Nazionale di Fisica Nucleare, Laboratori Nazionali di Legnaro, Legnaro, I-35020, Italy}
\address[IGFAE]{IGFAE, Universidade de Santiago de Compostela, E-15782, Santiago de Compostela, Spain}
\address[ific]{Instituto de Fisica Corpuscular, CSIC-Universidad de Valencia, E-46071 Valencia, Spain}
\address[brat]{Department of Nuclear Physics and Biophysics, Comenius University in Bratislava, 84248 Bratislava, Slovakia}
\address[york]{School of Physics, Electronics and Technology, University of York, North Yorkshire YO10 5DD, United Kingdom}
\address[wars]{Faculty of Physics, University of Warsaw, Warsaw, Poland}
\address[ithemba]{iThemba LABS, National Research Foundation, P.O. Box 722, Somerset West 7129, South Africa}
\address[surr]{Department of Physics, University of Surrey, Guildford GU2 7XH, United Kingdom}
\address[kul]{KU Leuven, Institut voor Kern- en Stralingsfysica, Celestijnenlaan 200D, 3001 Leuven, Belgium}
\address[nipne]{``Horia Hulubei'' National Institute of Physics and Nuclear Engineering (IFIN-HH), R-077125, Bucharest, Romania}
\address[aarh]{Department of Physics and Astronomy, Aarhus University, DK-8000 Aarhus C, Denmark}
\address[ijc]{IJCLab, CNRS/IN2P3, Université Paris-Saclay, F-91405 Orsay, France
}
\address[tenn]{Department of Physics and Astronomy, University of Tennessee, Knoxville, Tennessee 37996, USA}
\address[czech]{Institute of Experimental and Applied Physics, Czech Technical University in Prague, Husova 5, Prague, Czech Republic}
\address[lib]{Department of Physics, Oliver Lodge Laboratory, University of Liverpool, Liverpool, L69 7ZE, United Kingdom}
\address[ill]{Institut Laue-Langevin, 38042 Grenoble, France}
\address[stfc]{STFC Daresbury Laboratory, Daresbury, WA4 4AD, Warrington, United Kingdom}
\address[koln]{Institut für Kernphysik, Universität zu Köln, 50937 Cologne, Germany}

\address[west]{Department of Physics, University of the Western Cape, P/B X17 Bellville 7535, South Africa}
\address[ubc]{Department of Physics and Astronomy, University of British Columbia, Vancouver, British Columbia V6T 1Z4, Canada}

\fntext[dead]{Deceased.}
\fntext[argo]{Present address:Physics Division, Argonne National Laboratory, Argonne, Illinois 60439, USA}

\begin{abstract}


Near the doubly-magic nucleus \nuc{Ni}{78} ($Z=28$, $N=50$), there has been a decades-long debate on the existence of a long-lived isomer in \nuc{Cu}{76}. A recent mass measurement claimed to have settled the debate, by measuring the energy of the isomer and shedding light on the structure of the nucleus. In this work, we present new, more accurate, and precise values of the half-lives of the isomeric and ground states in \nuc{Cu}{76}. Our findings suggest that both states have very similar half-lives, in the 600-700 ms range, in disagreement with the literature values, implying that they cannot be differentiated by their decay curves. These results raise more questions than they answer, reopening the debate and showing that the structures in \nuc{Cu}{76} are still not fully understood. 

\end{abstract}




\end{frontmatter}




\section{Introduction}
\label{introduction}

Isomers are intriguing excited nuclear states with long half-lives (T$_{1/2}$), sometimes comparable or even longer than that of the ground state (g.s.) of their nucleus. The reasons for these long half-lives are diverse, such as large angular momentum differences between states or shape coexistence \cite{Walker2023}. There is a strong ongoing experimental effort to measure and understand the underlying nuclear structure effects that produce them, since they have a significant impact on other research fields and practical applications~\cite{Garg2023}. In medicine, for instance, \nuc{Tc}{99m} (T$_{1/2}=6$~h) is widely used as a $\gamma$-emitting radiotracer in scans of bone metastases, myocardial perfusion or thyroid~\cite{Papagiannopoulou2017}. Other practical uses that are currently being considered are nuclear batteries by isomer depletion~\cite{Terranova2022} or a nuclear laser using \nuc{Th}{229m}~\cite{Thirolf2019}. Isomers that impact reaction rates in stellar environments have been coined as ``astromers" (for isomers with astrophysical consequences), and recent works suggest they may play a fundamental role in nucleosynthesis~\cite{Misch2020}. Finally, some studies have suggested that superheavy elements, which are relevant for nuclear physics, atomic physics, and chemistry, might be more stable in isomeric states than in their g.s., thus enhancing the possibility of being created~\cite{Xu2004}.

The region around the doubly-magic $^{78}$Ni is rich in isomers. With protons near the closed $Z=28$ shell and neutrons filling the $g_{9/2}$ orbital between $N=40$ and $N=50$, there are ample opportunities for spin isomers to form. Isomers have already been observed in odd-odd Cu ($Z=29$) isotopes up to $^{72}$Cu~\cite{Grzywacz1998}. Neutron excitations across $N=40$ create negative parity states via the $\pi p_{3/2} \otimes \nu g_{9/2}$ coupling. Transitions from these states to the lower-lying positive-parity ones are hindered, leading to long lifetimes. There is currently no information on excited states in $^{74}$Cu~\cite{nndc} and just recently an isomer has been located in \nuc{Cu}{78}~\cite{Pedersen2023}.

Following this trend, an isomer should be expected in \nuc{Cu}{76}, and indeed it was reported by Winger \textit{et al.} with T$_{1/2}=1.27(30)$~s in a decay experiment performed over three decades ago~\cite{Winger1990}. However, subsequent laser spectroscopy, mass measurement, and decay spectroscopy experiments have not been able to replicate their result. Hence, the existence of this isomeric state in \nuc{Cu}{76} seemed dubious, as it was reported in the NUBASE2020 evaluation~\cite{Kondev2021}. Just recently, Canete and collaborators~\cite{Canete2024} were the first to provide firm evidence of the existence of this isomer by mass measurements. They used the evaluated half-lives to match the isomeric state as the one having $J=3$ and a short half-life, while the g.s. should have the longer one. Finally, they proposed the existence of an internal decay transition between the isomeric and the ground state and further elaborated on the possible nuclear structure of \nuc{Cu}{76}. 

In this Letter, we present results of two $\beta$-decay experiments that call into question the T$_{1/2}=1.27(30)$~s measured by Winger \textit{et al.}~\cite{Winger1990}. This, in turn, challenges the interpretation from the results of Canete \textit{et al.}~\cite{Canete2024}, demonstrating that the \nuc{Cu}{76} conundrum is far from being resolved.

\section{Prior experimental studies in \nuc{Cu}{76}}

\begin{figure}[t]
	\centering 
	\includegraphics[width=0.99\columnwidth]{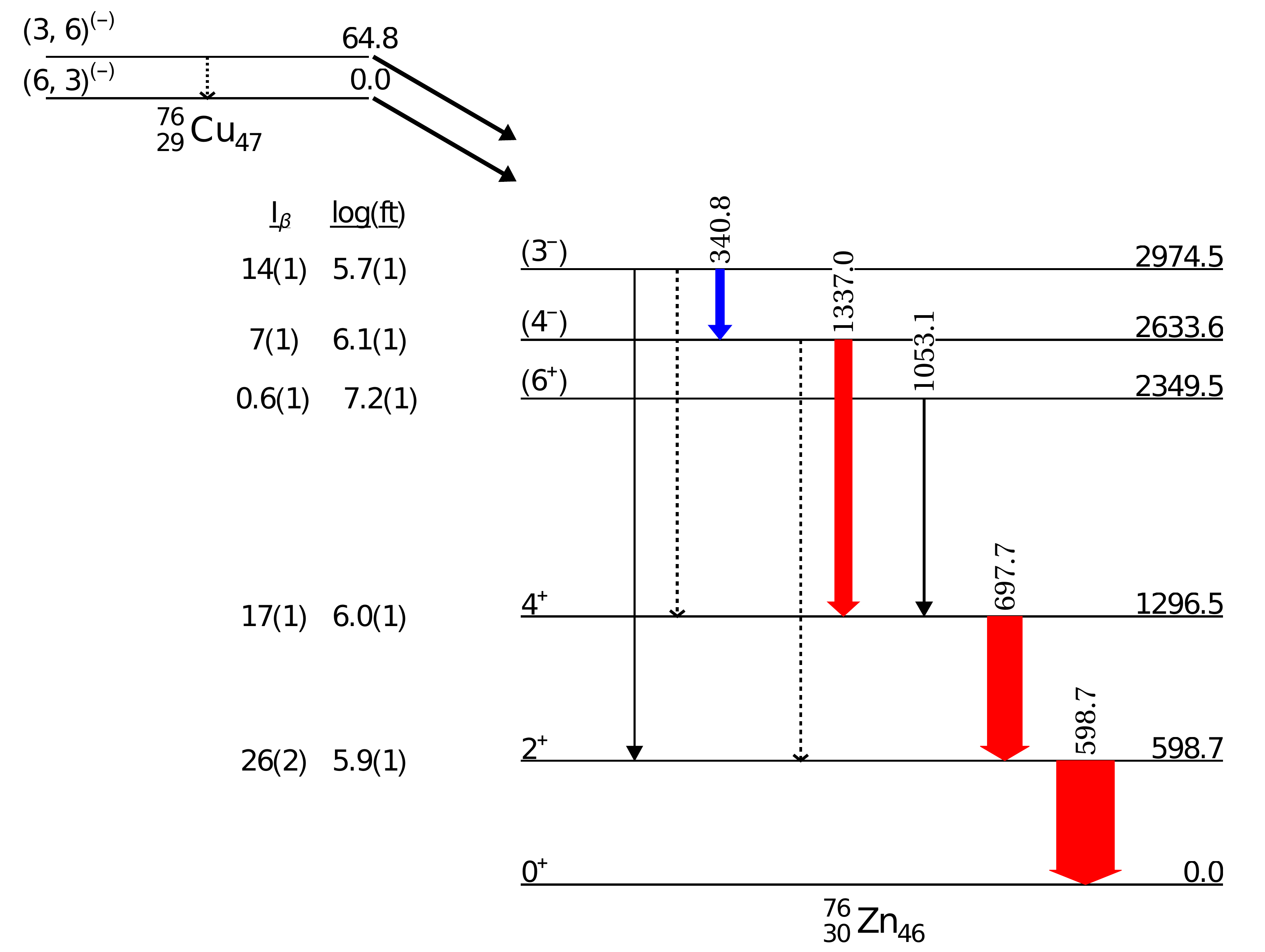}	
	\caption{The \nuc{Zn}{76} partial level scheme populated in the \nuc{Cu}{76} decay. For clarity, only the levels and transitions relevant to this work have been included. The internal transition in \nuc{Cu}{76} represented with a dashed line has never been observed.  The width of the lines is proportional to the intensity of the transition as observed in Exp.~I and their colour code follows the intensity scheme of NNDC. Similarly, the direct $\beta$-decay population ($I_\beta$) and the log(\textit{ft}) were extracted from Exp. I. The tentative spin-parities were deduced from the data of both experiments.
 }
	\label{Fig:level_scheme}%
\end{figure}

The existence of a $\beta$-decaying isomer in \cu{} was first suggested by Winger \textit{et al.}~\cite{Winger1990}. They produced \cu{} via fission by bombarding thermal neutrons into an uranium target at TRISTAN, in Brookhaven National Laboratory. They studied its decay to \nuc{Zn}{76} by gating on the \tran{4}{2} and \tran{2}{0} transitions, deducing half-lives of 0.84(6)~s and 0.57(6)~s, respectively. From that difference in half-lives, they inferred the existence of two $\beta$-decaying states in \nuc{Cu}{76} and, after subtracting the contribution from the shorter-lived one, they concluded that the other isomer had T$_{1/2}=1.27(30)$~s. They also proposed spins of $J \leq 3$ for the shorter-lived state and $J \geq 4$ for the longer-lived one based on the observed $\beta$-feeding pattern, (see an updated version of the level scheme in Fig.~\ref{Fig:level_scheme}). However, a number of the $\gamma$ rays that this experiment attributed to \nuc{Zn}{76} eventually were proven to belong to other decay products of the $A=76$ decay chain~\cite{Roosbroeck05}, suggesting the presence of unresolved contaminants. Furthermore, a period of just 1.4~s was used to record the decay curves, thus not being optimal to measure a T$_{1/2}=1.27(30)$~s value.

Since then, several experiments have studied the \nuc{Cu}{76} structure but have failed to observe the T$_{1/2}=1.27(30)$~s state~\cite{Kratz1991, Roosbroeck05, Guenaut2007, Winger2009, Hosmer2010, Welker2017, Groote2017, Silwal2022, Pedersen2024}. In these experiments, \nuc{Cu}{76} isotopes were created using a wide range of production techniques and all of them found half-lives compatible with $\sim 0.6$~s, but no trace of a longer-lived isomer. Below we discuss in detail some of these works.

Van Roosbroeck \textit{et al.}~\cite{Roosbroeck05} performed their experiment at ISOLDE using 1-GeV protons impinging on a thick UC$_x$ target. They expanded and corrected the \nuc{Zn}{76} level scheme proposed by Winger~\cite{Winger1990}. These corrections in the level scheme induced significant differences in the apparent $\beta$-decay feeding observed in the two experiments. 
In order to extract the \nuc{Cu}{76} half-life, they fitted the time evolution of the same two transitions as Winger \textit{et al.} Despite a similar amount of statistics, they observed no evidence of a longer-lived decaying state. It should be noted that the fitted time window only covered the range of 0-1.6~s, again sub-optimal to measure a half-life of 1.27~s.

Prior to the present work, Silwal and collaborators performed the largest statistics experiment on this decay~\cite{Silwal2022}. The production method was the bombardment of an UC$_{x}$ target by 54-MeV protons at Holifield Radioactive Ion Beam Facility (HRIBF), Oak Ridge National Laboratory (ORNL). The very high statistics allowed them to greatly expand the \nuc{Zn}{76} level scheme built by Van~Roosbroeck \textit{et al.}~\cite{Roosbroeck05}. Like Winger~\textit{et al.}~\cite{Winger1990}, Silwal~\textit{et al.} measured the \nuc{Cu}{76} half-life by gating on the \tran{4}{2} and \tran{2}{0} transitions. The results from both transitions were in agreement with each other and yielded an average of T$_{1/2} = 637(20)$~ms, ruling out T$_{1/2} \sim 1.27$~s. In this experiment, they implanted for 5 seconds and let it decay for 1~s. They simultaneously fitted both time periods, which although not ideal to measure a 1.27-s half-life, should have been able to observe it.

De Groote \textit{et al.} performed a collinear resonance ionization spectroscopy experiment at ISOLDE~\cite{Groote2017}. They generated neutrons by the impact of 1.4-GeV protons onto a neutron converter. Subsequently, the neutrons induced fission in a thick UC$_x$ target, producing Cu isotopes. At the time, they reported only one state in \nuc{Cu}{76} and firmly assigned $J=3$, which was hypothesized to be the ground state. 

Additionally, two mass-measurement experiments were performed at ISOLDE using a Penning trap~\cite{Guenaut2007, Welker2017}. Once again, fast neutrons were used to induce fission in an UC$_{x}$ target to produce the Cu isotopes. Both experiments only reported observing one state and their results were in perfect agreement between each other.

Up to this point in the story, it seemed that the existence of a second $\beta$-decaying state in \nuc{Cu}{76} was highly unlikely. In a turn of events, a very recent publication has presented the first firm evidence of its existence. Canete \textit{et al.} performed high-precision mass measurements at the Ion Guide Isotope Separator On-Line (IGISOL) facility~\cite{Canete2024, Giraud2022}. The \nuc{Cu}{76} isotopes were produced by 35-MeV protons impinging on an $^\text{nat}$U target. Using the time-of-flight ion-cyclotron-resonance (ToF-ICR) and phase-imaging ion-cyclotron-resonance (PI-ICR) techniques, they conclusively proved the existence of two long-lived states in \nuc{Cu}{76}, and they established that they are 64.8(25)~keV apart. By injecting the ions into the trap at three different times (389, 489, and 1089 ms), they attempted to measure the half-lives of both states. For what they established as the isomeric state, they obtained a T$_{1/2} = 672(110)$~ms, in agreement with the evaluated value of T$_{1/2} = 637(7)$~ms~\cite{Singh2024}. Since this was the half-life observed by most experiments previously, they assigned the $J=3$ measured by de~Groote~\cite{Groote2017} to this state. For what they established as the g.s., there is no statistically significant variation between the number of ions for the three measured times. The error bars for this latter state are too large and they could not perform a reliable fit. According to Canete and collaborators, this hints at a longer half-life and was the basis for assigning the g.s. as the T$_{1/2} = 1.27(30)$~s level, inverting the ordering of states previously suggested~\cite{Groote2017}. The authors explained their rather flat time distribution (see Fig. 3 in Ref.~\cite{Canete2024}) by assuming the existence of an internal decay between the isomer and ground states with a branching ratio of $10-17\%$. A limitation of this approach is that they only had three experimental points to fit a complex time distribution with the Bateman equations. 

Two even more recent experiments, whose data remain unpublished, were performed at RIKEN, which employs in-flight fragmentation, very different from all previous experiments, and that is expected to produce both states. The first experiment observed the \nuc{Cu}{76} decay using the BRIKEN setup~\cite{Tolosa2019} to measure $\beta$-$n$ branches and half-lives. As can be seen in Fig.~C.6 of the Ph.D. thesis~\cite{Tolosa2020}, the decay curve only presents one apparent half-life compatible with the evaluated T$_{1/2} = 637(7)$~ms. The second experiment was the first one to study the $\beta$ decay of \nuc{Ni}{76} to \nuc{Cu}{76}~\cite{Pedersen2024}. A decay from the $J^\pi=0^+$ g.s. of \nuc{Ni}{76} should (near-)exclusively feed into a low-spin isomer in \nuc{Cu}{76}. Despite their dedicated search for isomeric states in \nuc{Cu}{76}, they did not detect any $\sim 65$-keV $\gamma$ ray that would originate from its internal decay. The only hint of the isomer is that their fit to the lifetime of \nuc{Ni}{76} would slightly improve when introducing a second $\beta$-decaying state in \nuc{Cu}{76} with a longer half-life. However, they did not quantify this improvement, nor the required half-life. 

\section{Experimental details}

The data presented in this Letter were obtained in two experiments performed at ISOLDE, in two different campaigns but using the same production method and very similar setups. Protons were accelerated to 1.4 GeV and then impacted on a Ta proton-to-neutron converter. The neutrons induced fission on an UC$_{x}$ target and the Cu isotopes were ionized by Resonance Ionization Laser
Ion Source (RILIS) using a broad-band laser. Mass $A = 76$ was selected using a magnetic mass selector.

In both experiments, the experimental setup employed was the ISOLDE Decay Station (IDS). The ions were implanted on a moving tape at the center of the setup. In front of the moving tape, a fast plastic scintillator was installed to detect $\beta$ particles. To each side of it, two small LaBr$_3$(Ce) crystals coupled to fast photomultiplier tubes (PMT) were installed. During the first experiment (Exp. I), the setup consisted of 4 clover-type HPGe detectors, for a total of 16 germanium crystals. The data acquisition system employed was NuTAQ~\cite{Nutaq}. For further details of this setup, the reader is referred to Refs.~\cite{Benito2020,Piersa2021}. For the second experiment (Exp. II), the experimental setup was identical with the exception that two additional HPGe clovers were added downstream. Additionally, the DAQ was changed to Pixie16 by XIA. Exp. I was focused on the study of the lifetimes of excited states in \nuc{Zn}{76}, whereas Exp.~II was devoted to studying the decay of \nuc{Cu}{76} and its possible isomer. 


\section{Results}

\begin{figure}
	\centering 
	\includegraphics[width=\columnwidth]{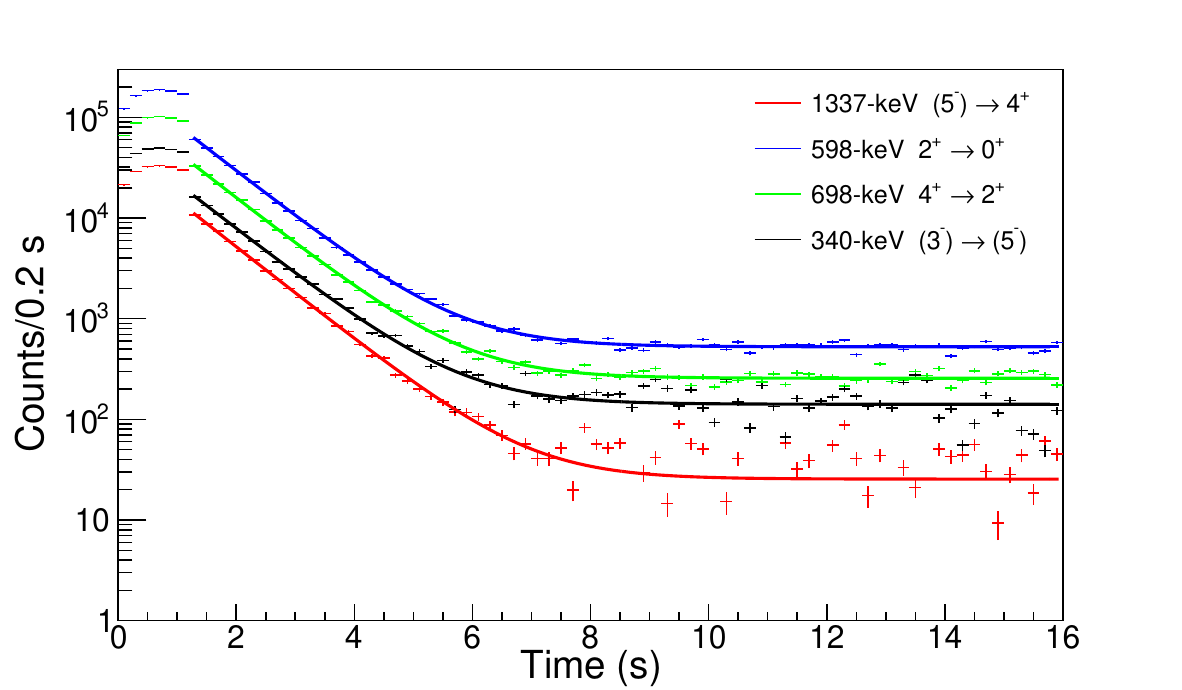}	
	\caption{The decay curve of \nuc{Cu}{76} was measured by gating in the four most intense $\gamma$-ray transitions in \nuc{Zn}{76} using data from Exp. II. The fit was done to an exponential decay plus a constant background, with good agreement between the four values. Some of the data points and curves were vertically displaced to avoid overlapping. See text for details.}
	\label{Fig:halflives}%
\end{figure}

The half-life of \nuc{Cu}{76} was extracted using the time of the proton beam impinging on the target as a reference. The four most intense $\gamma$-ray transitions in \nuc{Zn}{76}, which form a cascade (see Fig.~\ref{Fig:level_scheme}), were gated on to minimize contaminants and search for differences in decay rates. Figure~\ref{Fig:halflives} shows the four time distributions from Exp.~II and their fits to a single exponential decay plus a constant term to account for the time-random background. The Compton background was subtracted. Proton pulses arrive at ISOLDE in multiples of 1.2~s. For Exp.~II, a number of consecutive proton pulses would impinge on the target, followed by a ${\sim}30$~s decay time. Since the employed time reference resets to zero every time a proton bunch arrives at ISOLDE, this explains the step observed at 1.2~s in the decay curves. As it can be observed, all four curves follow the same decay rate. The apparent deviations at later times are due to differences in the peak-to-background ratio. The results are shown in Tab.~\ref{Tab:Halflives} and yield an average half-life of T$_{1/2}=656(2)$~ms for Exp. II. The results from Exp.~I are deemed of lower quality for this specific observable and are just presented to show that they are compatible. The decaying rate of the 1053-keV \tran{(6)}{4} transition was also fitted, yielding T$_{1/2}=657(22)$~ms, in agreement with the other values, although with a much larger error bar due to its lower statistics. For this reason, it was not included in Fig.~\ref{Fig:halflives}.

It should be highlighted here that these results are not only more precise due to the much higher level of statistics accumulated, but they are potentially more accurate due to the significantly longer decay time. On top of performing the fit to a time range of several lifetimes, this allows the fitting of the background, which has a non-negligible influence on the result. None of the previous experiments could accurately assess this.


\begin{figure}[t]
	\centering 
	\includegraphics[width=\columnwidth]{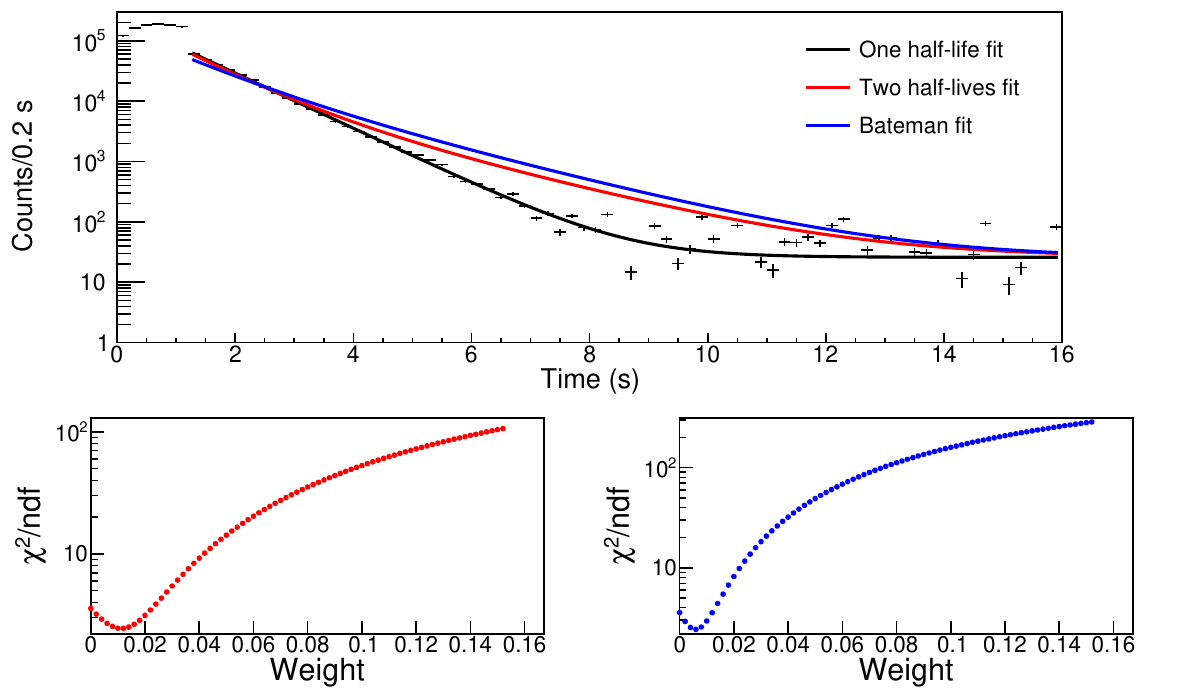}	
	\caption{Two fits to the \nuc{Cu}{76} decay when gating on the 598-keV \tran{2}{0} transition in \nuc{Cu}{76}. In black, the fit to a single exponential decay plus a constant background. In red, fit to two independent decays with a fixed component of $10\%$ and T$_{1/2}=1.27$~s. In blue, same as red, but the second decay component is instead a Bateman equation to simulate the internal decay between isomer and ground state in \nuc{Cu}{76}. The two lower panels show the $\chi^2$ of the fit as a function of the weight of the 1.27-s component. In both cases, an upper limit of $<2\%$ can be set for such a long-lived state. Data from Exp II.}
	\label{Fig:bateman}%
\end{figure}

To study a possible feeding from a T$_{1/2}=1.27(30)$~s decay in \nuc{Cu}{76}, additional fits to the time distribution of the 598.7-keV \tran{2}{0} transition were performed. Given that it is the lowest-lying transition, it is expected to collect the intensity from the transitions populated by both, the ground and isomer states in \nuc{Cu}{76}, as suggested by Winger~\cite{Winger1990}. The fit was performed assuming two decaying states, one with a fixed half-life of T$_{1/2}=1.27$~s and the other with a free T$_{1/2}$. Figure~\ref{Fig:bateman} shows the result when the ratio between the two components was fixed as $10\%$ for the longer T$_{1/2}$ and $90\%$ for the free one. The red fit simulates that both states in \nuc{Cu}{76} decay via $\beta^-$ emission independently. The blue fit makes use of the Bateman equations to simulate an internal decay with a branching ratio of $10\%$, plus a 90\% direct $\beta^-$ emission. In both cases, the normalization factor (related to the initial number of nuclei) and the constant background remained as free parameters. The black fit is a single exponential decay with a free half-life, same as Fig.~\ref{Fig:halflives}, and it is shown for reference. The fits containing two half-lives clearly deviate from the experimental data, ruling out the presence of a T$_{1/2}=1.27$~s at the $10\%$ level, as suggested by Canete~\cite{Canete2024} and Winger~\cite{Winger1990}. In the two lower panels, the $\chi^2$ of the fits were studied as a function of the proportion of the 1.27-s component [be it as two independent decays (red, left panel) or via an internal decay following the Bateman equation (blue, right panel)]. These set an upper limit of $<2\%$ for either the direct production of the T$_{1/2}=1.27(30)$~s state or of an internal decay to it. 

\begin{table*}
\centering
\begin{tabular}{c c c c c c c c c} 
 \hline
Transition	&	NNDC adopted		&	Exp. I		&	Exp. II		&	Silwal		&	Hosmer		&	Van Roosbroeck		&	Kratz		&	Winger		\\
(keV)	&	\cite{Singh2024}		&			&			&	\cite{Silwal2022}		&	\cite{Hosmer2010}		&	\cite{Roosbroeck05}		&	\cite{Kratz1991}		&	\cite{Winger1990}		\\ \hline
598.7	&			&	   665(9)     &657(2)				&	632(21)	&			&			&			&	1270(30)	\\
697.7	&			&	 668(14)       &656(2)				&	640(33)	&			&			&			&	570(60)	\\
1337.0	&			&	    647(21)    &651(4)				&			&			&			&			&			\\
340.8	&			&	    670(21)    &661(5)				&			&			&			&			&			\\ \hline
Total	&	637(7)	&	   664(7)     &656(2)				&	637(20)	&	599(108$^*$)	&	653(24)	&	641(6)	&	570(60)	\\

 \hline
\end{tabular}
\caption{Comparison of the measured half-lives in different decay experiments. When the half-life was deduced by gating on a specific transition, it is placed in the row of that transition. Final average values and results for which the information on gating was not given, are shown in the row "Total". All half-lives are in ms. $^*$~This is a corrected value of the uncertainty reported in Ref.~\cite{Hosmer2010}, see main text for details.}
\label{Tab:Halflives}
\end{table*}

\begin{figure}
	\centering 
	\includegraphics[width=\columnwidth]{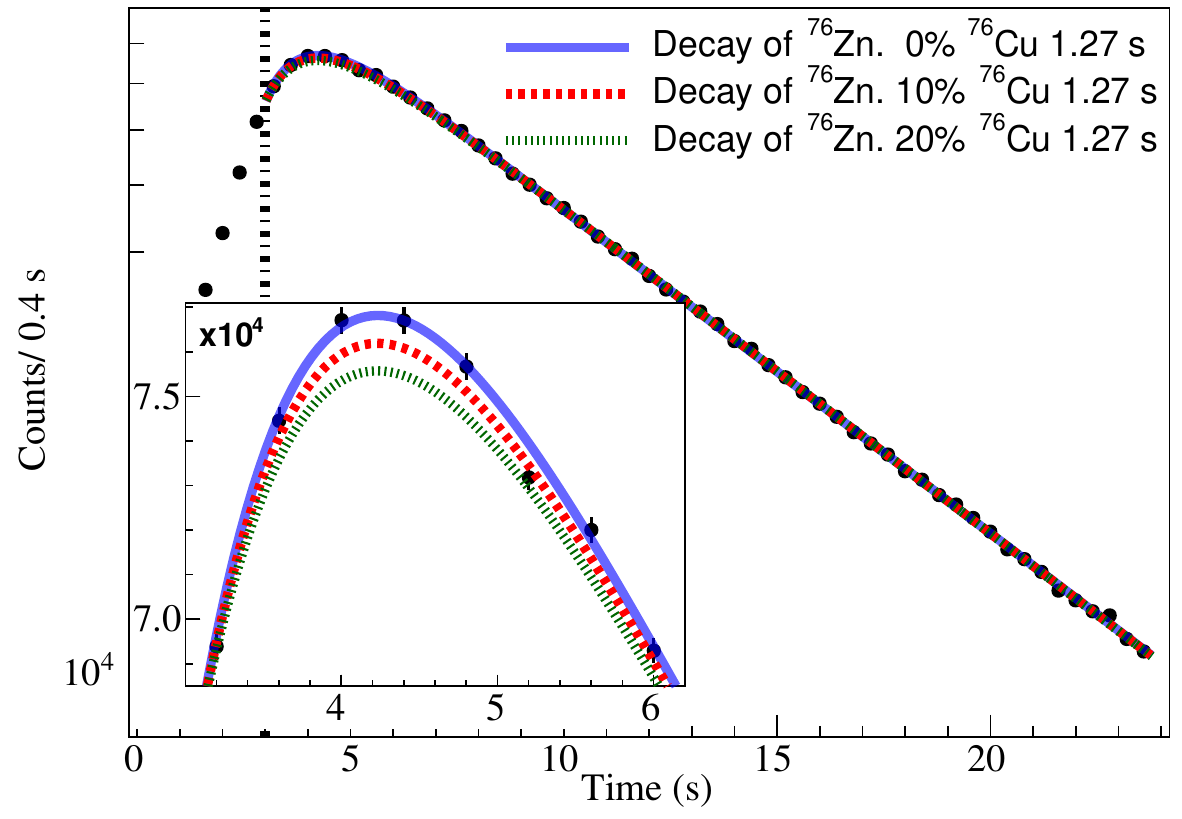}	
	\caption{Fit to the \nuc{Zn}{76} decay when selecting the 199-keV $1^+ \rightarrow 2^{(-)}$ transition in \nuc{Ga}{76}. The black dots represent the experimental data. The vertical black-dashed line represents the closing of the beam gate. The colour lines represent the fits to the whole distribution. In these functions, two contributions were included. First, the decay of \nuc{Zn}{76} that is already present when the beam gate is closed. The second contribution accounts for the later population of \nuc{Zn}{76} produced by the decay of \nuc{Cu}{76} after the beam-gate closure. The blue, red, and green curves correspond to the cases when a 0\%, 10\%, and 20\% of \nuc{Cu}{76} with T$_{1/2}$~=~1.27(30) s are considered, respectively. The inset shows a zoom in the area of the maximum activity, a few seconds after the closure of the beam gate. This region corresponds to the area where the influence of the \nuc{Cu}{76} activity is most relevant. Data from Exp II.}
	\label{Fig:Zn_decay}%
\end{figure}

The possibility that the T$_{1/2}$~=~1.27(30)~s \nuc{Cu}{76} decay populates mostly the g.s., preventing its observation in the time distribution of \nuc{Zn}{76} $\gamma$-rays, was also studied. To investigate this hypothetical scenario, $\gamma$-ray transitions in the \nuc{Ga}{76} granddaughter were analyzed. Figure~\ref{Fig:Zn_decay} shows the time distribution of the 199-keV $\gamma$-ray, which is the most intense transition in \nuc{Ga}{76}. After the closure of the beam gate, the implantation period is stopped, therefore the time distribution can be described as the sum of two components. The first component corresponds to the decay of the \nuc{Zn}{76} already present when the beam-gate is closed and thus, it has an exponential decay with the half-life of \nuc{Zn}{76}. The second one arises from the decay of \nuc{Cu}{76} after the closure, causing an increase in the activity whose shape depends on the half-life of \nuc{Cu}{76}. Therefore, the shape of the decay curve of the 199-keV transition is described by the Bateman equations. The fitting results of the time distribution were obtained by fixing the half-lives and setting the normalization factors as free parameters. In this case, only the contribution of the 656(2)~ms state (blue-line curve) was assumed, and hence, an excellent match to the time distribution was achieved, see Fig.~\ref{Fig:Zn_decay}. On the other hand, when assuming that 10\% or 20\% of \nuc{Cu}{76} belongs to a possible feeding of the 1.27-s component the fit deteriorates significantly, see the red and green lines in the same figure. This is highlighted in the inset of Fig.~\ref{Fig:Zn_decay}, where the fits with a $10\%$ or $20\%$ component clearly show a deviation at the 4-s time mark. The analysis yielded an upper limit of 5\% for the 1.27-s component, if any. This result supports the conclusion derived from $\gamma$ rays in the previous analysis, excluding the existence of any significant T$_{1/2}=1.27(30)$~s decay. It should be noted that during the analysis of these data, a half-life for \nuc{Zn}{76} of 6.44(4)~s was measured. This value significantly deviates from the adopted T$_{1/2}$~=~5.7(3)~s~\cite{Singh2024}. However, the adopted value was obtained from the integral of the $\beta$(t)~\cite{GRAPENGIESSER1974} without any $\gamma$-ray selectivity.

\begin{figure}
	\centering 
	\includegraphics[width=\columnwidth]{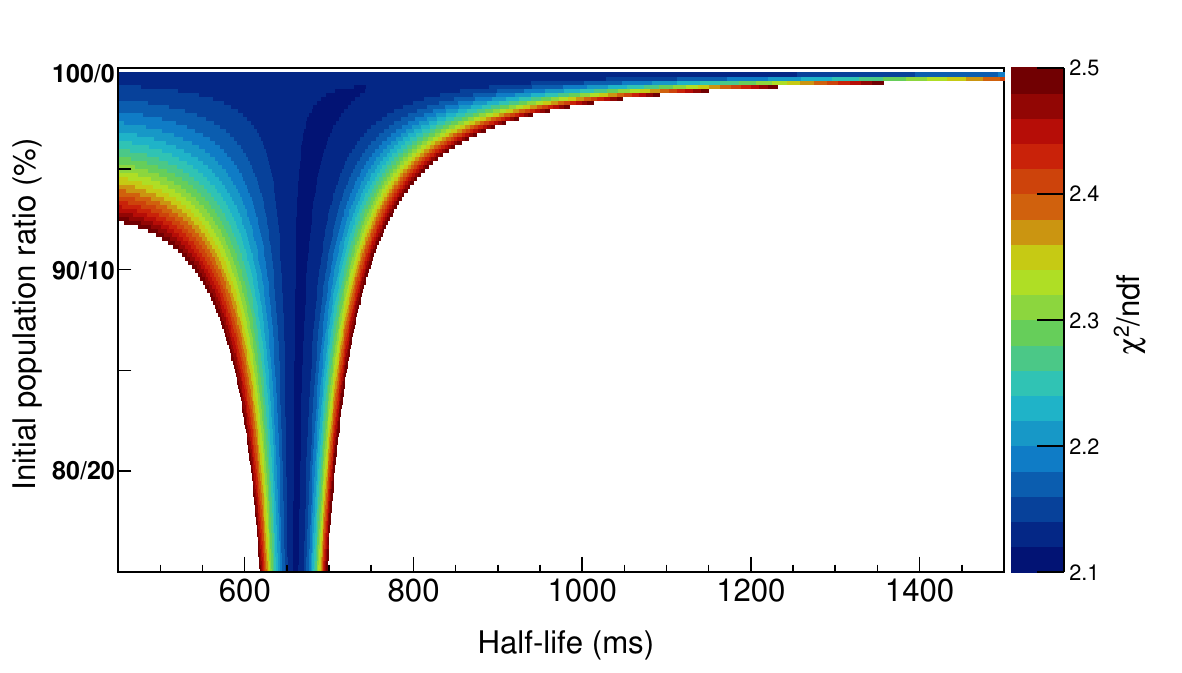}	
	\caption{ Fit to the time profile of the \tran{2}{0} transition in \nuc{Zn}{76} to two independent half-lives. One was fixed to T$_{1/2}$~=~656~ms while the other and the ratio between both components was free. The x-axis represents the free half-life in ms. The y-axis shows the percentage initial population of both decaying states. The top of the y-axis represents that 100\% of the population is that of the free half-life component and at the bottom, 75\% is the free half-life and 25\% the fixed one. The z-axis shows the normalized $\chi^2$ of the fit.}
	\label{Fig:lifetimes_map}%
\end{figure}

The T$_{1/2}$~=~1.27(30)~s value reported by Winger has a large error bar~\cite{Winger1990}, so the actual half-life of the decaying state might be significantly lower. A fit to the activity versus time of the \tran{2}{0} transition in \nuc{Zn}{76} was performed, similarly to Fig.~\ref{Fig:bateman}. This time one of the lifetimes was fixed to T$_{1/2}$~=~656~ms, while the other one and the ratio between both components was free. Figure~\ref{Fig:lifetimes_map} shows a colour map with the result, see its caption for details. The result presents a sharp minimum along the line of both half-lives having very similar values. The free component has an upper limit of about 970~ms when its weight reaches $\sim 1.5\%$, which is the lower one-sigma bound of the value reported by Winger. When the free component of the half-life is lower than the fixed T$_{1/2}$~=~656~ms, the increase in $\chi^2$ seems to be less steep. This possibility is discussed, and discarded, in depth in the discussion section. 

\section{Discussion}

\subsection{Are both \nuc{Cu}{76} decaying states produced?}

Before any discussion can proceed, whether or not \nuc{Cu}{76} is being produced in two different long-lived states needs to be established. Canete \textit{et al.}~\cite{Canete2024} have conclusively proven that that is the case at IGISOL, when using 35-MeV protons on a U thin target. It is known that isomeric ratios can change for different fissions systems or projectile energies, but it is a modest change, not an order of magnitude. In fact, at ISOL facilities that employ thick targets, the release speed plays a more dominant role. This is the case for Cu isotopes, that can take up to a few seconds to exit a thick target as the ones employed at ISOLDE~\cite{Koester2000}. Short-lived species will decay before they can diffuse out of the target, tilting the ratio towards the longer-lived isomer. Compounding these two effects, it could be expected that the isomeric ratio between IGISOL and ISOLDE to be of the same order of magnitude.

A strong hint of this being the case arises from the two mass measurements performed at ISOLTRAP~\cite{Guenaut2007, Welker2017}. Both experiments yielded masses that are within one standard deviation from each other, indicating that the results are consistent. However, when compared with the ones obtained at IGISOL~\cite{Canete2024}, the ISOLDE results are 29.8(22)~keV higher than the g.s. and around halfway to the excitation energy of the isomer state. The most likely explanation for this situation is that the experiments were not able to separate their masses, yielding an intermediate result. This interpretation was already suggested in Ref.~\cite{Canete2024}. The fact that the experiments yielded an average of the masses of the two states is a known feature of mass measurements for unresolved species~\cite{Hukkanen2023}. 

Another indication of the production of both \nuc{Cu}{76} states at ISOLDE comes from a reexamination of the laser spectroscopy experiment performed by de Groote~\cite{Groote2017}. A new analysis of the data presented in their Fig.1, top panel, showed hints of a $J=6$ state (see the excess counts above background around $-370$ and $50$~MHz), compatible with a production at the level of a few \% for this higher-spin state~\cite{GrootePC}. This is a tentative result that requires of a dedicated laser spectroscopy experiment to exhaustively scan an appropriate frequency range to conclusively identify this state and assign it as $J=6$. 

A more general hint, not specific to ISOLDE, arises from the $\beta$-feeding pattern of states in \nuc{Zn}{76}. All decay experiments so far have observed a significant direct $\beta$-decay population of the main four states shown in the partial level scheme of Fig.~\ref{Fig:level_scheme}. The first two levels have been firmly established as the $2_1^+$ and $4_1^+$ states~\cite{Singh2024}. There are different suggestions for the spins and parities ($J^\pi$) of the other two levels, although the consensus seems to be that they belong to a negative-parity multiplet \{3, 4, 5, 6\} originating from a $\pi p_{3/2}\otimes\nu g_{9/2}$ coupling. Especially relevant is the state at around 2.6 MeV, which strongly feeds the $4_1^+$ state but does not feed the g.s., and probably does not feed the $2_1^+$ state. This suggests a $J^\pi\geq4^-$ assignment. Although not impossible, it is extremely unlikely that a single $J^\pi=3^-$ state in \nuc{Cu}{76} simultaneously populates with log(\textit{ft})${\sim}6$ states with spins $2^+$ and $4^+$ (which would be of first forbidden character) and states with spins $3^-$ or $4^-$ (which would be of allowed character)~\cite{Turkat2023}. If any of the negative-parity states have $J\geq 5$, as has been suggested, the forbiddenness of the transition would be of second or higher order, making this direct transition even more unlikely. A much more plausible explanation is that the decay of \nuc{Cu}{76} originates from two different states. 

Although all experiments have observed strong population to the four states discussed in the previous paragraph, the exact $\beta$ intensities vary from work to work, which could be an indication of different isomeric ratios. However, comparing $\beta$-decay feedings between different experiments is troublesome. Experiments that record different amount of statistics will measure a different number of $\gamma$-ray transitions. This, in turn, can have a significant influence on the apparent $\beta$ feeding, which is known as the \textit{Pandemonium} effect~\cite{Hardy1977}. Moreover, due to the complexity of building a level scheme with dozens or hundreds of transitions, it is not uncommon that they contain small errors, such as overlooked transitions or wrongly placed ones. This compounds with the previous effects to further complicate the comparison between different experiments.

To avoid these issues, instead of comparing $\beta$ feeding between the different experiments, the relative intensities of the four most intense transitions are compared, see Tab.~\ref{Tab:Intensities}. The 2350-keV state has been suggested to have $J^\pi = (6^+)$~\cite{WaltersPC, Silwal2022}, which is the highest spin proposed for a level in this nucleus. Therefore, the 1053-keV \tran{(6)}{4} transition is a candidate to observe population from the higher-spin decaying state in \nuc{Cu}{76} and has been included in the table despite its lower intensity. As can be seen, there is a good agreement among Exp. I and Exp. II from this work and the results of Van Roosbroeck~\cite{Roosbroeck05}, also performed at ISOLDE. However, those of Silwal~\cite{Silwal2022} that employed a different fission system and beam energy, are more than four standard deviations away. Such a discrepancy may be due to a different isomeric ratio at that facility compared to ISOLDE.

\begin{table}[b]
\centering
\begin{tabular}{c c c c c c} 
 \hline
Transition	&	Exp. I		&	Exp. II		&	Silwal		&	Van Roosbroeck		\\
(keV)	&			&			&	\cite{Silwal2022}		&	\cite{Roosbroeck05}		\\ \hline
598.7	&	100(2)	&	100(2)	&  100.0(21)   &  100(5)	\\
697.7	&	60(1)	&	59.5(6)	&	66.6(11)   &	59(3)	\\
1337.0	&	33.1(6)	&	32.8(6)	&	38.3(7)	   &	30(2)	\\
340.8	&	17.0(3)	&	19.1(2)	&	19.5(4)	   &	16(1)	\\
1053.1	&	3.82(8)	&	 4.0(1)	&	 5.06(17)  &	 3.1(9)	\\
 \hline
\end{tabular}
\caption{Relative intensities of the four most intense $\gamma$-ray transitions in \nuc{Zn}{76} following the decay of \nuc{Cu}{76}. The 1053.1-keV \tran{(6)}{4} transition has also been included as a clear example of a transition decaying from a high-spin state [$J^\pi=(6^+$)]. Exp. I and II refer to the results from this work. In all experiments, the intensities have been normalized for the 598.7-keV \tran{2}{0} transition to have an intensity of 100. As can be seen, the results of Silwal~\textit{et al.}~\cite{Silwal2022} significantly deviate from the other experiments. See text for details.
}
\label{Tab:Intensities}
\end{table}

While IGISOL is the only facility that has firmly established the production of both the ground and isomer states, we believe that the previously discussed points prove that both states are produced in other laboratories as well, particularly at ISOLDE.


\subsection{The different \nuc{Cu}{76} half-lives}

Once established that there is an isomer in \nuc{Cu}{76}~\cite{Canete2024} and that laboratories produce \nuc{Cu}{76} in its two $\beta$-decaying states, it raises the question as to why no other experiment has observed the T$_{1/2}=1.27(30)$~s proposed by Winger \textit{et al.}~\cite{Winger1990}. As can be seen in Tab.~\ref{Tab:Halflives}, all half-lives but that obtained by Hosmer~\cite{Hosmer2010}, agree on a value of $T_{1/2}=630-660$~ms. Hosmer \textit{et al.} performed their experiment at the NSCL in-flight fragmentation facility, using a thin target, which should be more sensitive to shorter-lived isomers when compared to facilities employing thick targets such as ISOLDE or HRIBF-ORNL. While the reported half-life is indeed shorter, we note that for the values of \nuc{Cu}{76} and \nuc{Ga}{81} in column 3 of their Table I, there was an accidental mix-up between relative uncertainty (\%) and absolute uncertainty (ms), as evident from the low number of implanted ions. In reality, the relative uncertainty of the \nuc{Cu}{76} half-life is 18\%, or $\sim100$~ms absolute uncertainty. Therefore, the value from Ref.~\cite{Hosmer2010} provides insufficient constraints on the half-life and will be disregarded in the further discussion.  We note that the value of Hosmer and its erroneously small uncertainty also entered the weighted average providing the half-life adopted by ENSDF~\cite{Singh2024}.


The most likely scenario is that the half-lives of both decaying states are in the 600-700~ms range. It is nearly impossible to disentangle two half-lives in a mixed decay when their values are so similar. There are multiple instances of this phenomenon throughout the nuclear chart (see a few examples of such nuclei in Refs.~\cite{Hagberg1997, Pauwels2009, Cerny1972, Roosbroeck2004, Paziy2020}). 

The only way to discriminate between such similar half-lives would be by identifying states that are only directly fed by one of the decays and not indirectly fed by the other one. While it is likely that some of the levels at higher energy fulfil these conditions, none of their transitions have high-enough intensity to perform the analysis with sufficient precision.

The other possible scenario in which one of the decaying states in \nuc{Cu}{76} has a very short half-life can be excluded. The fact that the mass measurement experiments seem to observe both states at late times~\cite{Guenaut2007, Welker2017,Canete2024}, excludes the possibility of half-lives below a few hundred of ms.


\subsection{The $J=3$ assignment to the isomeric state and the internal decay branch}

Excluding the existence of a $1.27(30)$~s half-life in \nuc{Cu}{76} calls into question some of the conclusions reached by Canete \textit{et al.}~\cite{Canete2024}. As explained earlier, from the observed time distributions, they assigned the T$_{1/2}=1.27(30)$~s value to the g.s. and the $J=3$ spin measured by de Groote~\cite{Groote2017} to the isomer, which was an inversion to what had been previously proposed. However, now that it has been proven that both states have similar half-lives, these assignments are challenged. If the half-lives of both states are so close that they cannot be easily differentiated, the time distributions observed by Canete \textit{et al.} cannot be used to assign spins. Therefore, which of the two states has the $J=3$ spin remains unanswered. 

This has further implications. In their work, Canete \textit{et al.}~\cite{Canete2024} assumed that most \nuc{Cu}{76} was produced in the T$_{1/2}\sim 650$~ms isomeric state and a $10-17\%$ internal decay branch populated the T$_{1/2}=1.27(30)$~s g.s. They claimed this was necessary to explain the flatness of the observed time distribution, which they fitted using the Bateman equations. Given that every decay experiment observed a ${\sim}0.65$~s decay, under this scenario \nuc{Cu}{76} would be mainly produced in its isomeric state in every laboratory. The subsequent $10-17\%$ internal decay branch would populate the g.s. and every decay experiment should have observed a longer-lived component in their decay curves. Figure~\ref{Fig:bateman} excludes this scenario, when even a $10\%$ component of the T$_{1/2}=1.27(30)$~s strongly distorts the fit.

There is an additional nuance. The internal decay suggested by Canete~\cite{Canete2024} would proceed by a low-energy 64.8-keV $\gamma$ transition, which would be strongly converted. The emission of a conversion electron would increase the charge state of the ion to Cu$^{2+}$ (or higher considering subsequent emission of Auger electrons). This would change the q/m of the ions and not be counted by the device, i.e. the actual branching ratio of the internal decay would be significantly higher. For instance, the conversion coefficient for an E2 transition is 3.17 and for M3 it is 39~\cite{Kibedi2008}. This translates the 10\% lower bound of the branching ratio to 37\% and 98\%, respectively. Such intense decay branches would have been observed by these and previous experiments, which is not the case.



Returning to the long half-life fit reported by Canete \textit{et al.}~\cite{Canete2024}, it is clear that the authors were limited by the experimental conditions. Their fitting conditions, i.e. the time range, the number of data points, and the technique, were not optimal for this kind of measurements, making this endeavor very challenging. It might be possible to repeat now the analysis using similar half-lives of T$_{1/2}\sim 650$~ms for both states. However, the change induced in the effective half-life when compounding two similar lifetimes in the decaying part of the Bateman equations is nearly negligible. Even the high-statistics experiments presented here did not attain enough precision to differentiate such an effect. Given the current experimental knowledge, the existence of this internal decay branch cannot be confirmed or refuted, but it is currently not needed to explain the experimental observations.


\subsection{Possible $J^\pi$ assignments of the long-lived states}

Tentative spin and parity assignments may be explored for the long-lived states in \nuc{Cu}{76}. One of  these states has been firmly assigned $J=3$ based on laser spectroscopy~\cite{Groote2017}. However, there is no experimental evidence to determine its parity. Nevertheless, based on systematics of even-Cu isotopes and shell-model calculations, it is likely to have negative parity and will therefore be considered $J^\pi=3^{(-)}$. The $\beta$-decay pattern in Fig.~\ref{Fig:level_scheme} shows a strong population of the 2974.5-keV $J^\pi=(3^-)$ state in \nuc{Zn}{76}, and a slightly weaker one to the $2^+_1$ and $4^+_1$ states, which can be explained as allowed and first-forbidden transitions from the parent $J^\pi=3^{(-)}$. For the 2633.6-keV state, the shell-model calculations in Ref.~\cite{Chester21} seem to favour $J^\pi=5^-, 6^-$ over the $(4^-)$ suggested in NNDC~\cite{nndc}. In this case, it could be directly populated, as the data show, if the other $\beta$-decaying state in \nuc{Cu}{76} also has $J^\pi=6^-$. This high-spin would also explain its isomeric, $\beta$-decaying character, as an M3 transition of 65 keV would be significantly hindered.

As mentioned earlier, there are tentative indications of a $J=6$ state in the laser spectroscopy~\cite{GrootePC}. This high-spin assignment is further supported by the $^{76}$Ni $\rightarrow$ Cu$^{76}$ experiment by Pedersen~\cite{Pedersen2019}. A decay from a $0^+$ state would be greatly suppressed to even indirectly populate $J^\pi=6^-$ states. The Monte Carlo shell-model calculations using the A3DA-m interaction from that work also rule out any low-lying positive-parity state in \nuc{Cu}{76}, discarding $J^\pi=1^+$, for example. Therefore, we propose $J^\pi=3^{(-)}$ and $J^\pi=(6^-)$ for the two decaying states in \nuc{Cu}{76}, in either order. It should be emphasized, however, that these assignments are highly tentative and that more experiments are needed for a firm assignment.

\section{Conclusions}

We have performed two high-statistics experiments to study the $\beta^-$ decay of \nuc{Cu}{76}. The results presented here, especially when added to the experimental corpus accumulated so far, clearly exclude the existence of a T$_{1/2}=1.27(30)$~s half-life in \nuc{Cu}{76}. While the existence of two long-lived states was firmly established by Canete \textit{et al.}~\cite{Canete2024}, the exact half-lives of these levels remain a mystery. After demonstrating that both, the isomer and ground state, are produced in the experiments, we propose that the most likely explanation for the current situation is that the two states have similar half-lives in the 600 to 700~ms range, making their separation significantly more difficult.

This change in the half-lives, in turn, questions some of the conclusions reached by Canete \textit{et al.}~\cite{Canete2024}. Which state has $J=3$ cannot currently be affirmed and the existence of an internal decay branch between them is highly unlikely. Experiments using isomerically pure beams could elucidate the situation, which, for now, remains unsolved.

\section*{Acknowledgements}

This work was supported by the European Union's Horizon 2020 research and innovation programme under grant agreement No. 654002 (ENSAR2). BO acknowledges funding from the RYC2021-031494-I and PID2022-140162NB-I00 projects. AI acknowledges funding from the European Union's Horizon 2020 research and innovation programme under the Marie Skłodowska-Curie grant agreement No. 847635. JB also acknowledges the support from the Margarita Salas Fellowship, CT31/21, at the Complutense University of Madrid funded by the Spanish MIU and European Union-Next-Generation funds. This work was supported by Spanish MCIN/AEI/10.13039/501100011033 under grant PID2021-126998OB-I00 and by Grupo de F\'isica Nuclear-UCM Ref.~910059, Ministerio de Ciencia e Innovación PID2022-138297NB-C21, Comunitat Valenciana Prometeo CIPROM/2022/9 and Severo Ochoa 
CEX2023-001292-S grants, by the Polish National Science Center under Grant No. 2020/39/B/ST2/02346, by the Polish Ministry of Science and Higher Education under contract 2021/WK/07, by the German BMBF under contract 05P21PKCI1 and Verbundprojekt 05P2021, by the Slovak Research and Development Agency (Contract No. APVV-22-0282), by the Slovak grant agency VEGA (Contract No. 1/0651/21), by the Academy of Finland project No. 354968, by the United Kingdom Science and Technology Facilities Council through the grants ST/P004598/1, ST/V001027/1, and ST/V001035/1, by the Romanian IFA grant CERN/ISOLDE and Nucleu project No. PN 23 21 01 02, and by the Research Foundation Flanders (FWO, Belgium), the BOF KU Leuven (C14/22/104), and the BOF KU Leuven (GOA/2015/010).

\appendix

\bibliographystyle{elsarticle-num} 
\bibliography{bibliography}

\def\url#1{}
\begin{thebibliography}{10}
\expandafter\ifx\csname url\endcsname\relax
  \def\url#1{\texttt{#1}}\fi
\expandafter\ifx\csname urlprefix\endcsname\relax\def\urlprefix{URL }\fi
\expandafter\ifx\csname href\endcsname\relax
  \def\href#1#2{#2} \def\path#1{#1}\fi

\bibitem{Walker2023}
P.~M. Walker, Z.~Podoly{\'a}k,
  \href{https://doi.org/10.1007/978-981-19-6345-2_46}{Nuclear Isomers},
  Springer Nature Singapore, Singapore, 2023, pp. 487--523.
\newblock \href {https://doi.org/10.1007/978-981-19-6345-2_46}
  {\path{doi:10.1007/978-981-19-6345-2_46}}.
\newline\urlprefix\url{https://doi.org/10.1007/978-981-19-6345-2_46}

\bibitem{Garg2023}
S.~Garg, B.~Maheshwari, B.~Singh, Y.~Sun, A.~Goel, A.~K. Jain,
  \href{https://www.sciencedirect.com/science/article/pii/S0092640X22000468}{Atlas
  of nuclear isomers—second edition}, At. Data Nucl. Data Tables 150 (2023)
  101546.
\newblock \href {https://doi.org/https://doi.org/10.1016/j.adt.2022.101546}
  {\path{doi:https://doi.org/10.1016/j.adt.2022.101546}}.
\newline\urlprefix\url{https://www.sciencedirect.com/science/article/pii/S0092640X22000468}

\bibitem{Papagiannopoulou2017}
D.~Papagiannopoulou,
  \href{https://pubmed.ncbi.nlm.nih.gov/28618064/}{Technetium-99m
  radiochemistry for pharmaceutical applications}, Journal of Labelled
  Compounds and Radiopharmaceuticals 60~(11) (2017) 502--520.
\newblock \href {https://doi.org/https://doi.org/10.1002/jlcr.3531}
  {\path{doi:https://doi.org/10.1002/jlcr.3531}}.
\newline\urlprefix\url{https://pubmed.ncbi.nlm.nih.gov/28618064/}

\bibitem{Terranova2022}
M.~L. Terranova,
  \href{https://onlinelibrary.wiley.com/doi/abs/10.1002/er.8539}{Nuclear
  batteries: Current context and near-term expectations}, International Journal
  of Energy Research 46~(14) (2022) 19368--19393.
\newblock \href
  {http://arxiv.org/abs/https://onlinelibrary.wiley.com/doi/pdf/10.1002/er.8539}
  {\path{arXiv:https://onlinelibrary.wiley.com/doi/pdf/10.1002/er.8539}}, \href
  {https://doi.org/https://doi.org/10.1002/er.8539}
  {\path{doi:https://doi.org/10.1002/er.8539}}.
\newline\urlprefix\url{https://onlinelibrary.wiley.com/doi/abs/10.1002/er.8539}

\bibitem{Thirolf2019}
P.~Thirolf, B.~Seiferle, L.~Von~der Wense,
  \href{https://iopscience.iop.org/article/10.1088/1361-6455/ab29b8}{The
  229-thorium isomer: doorway to the road from the atomic clock to the nuclear
  clock}, Journal of Physics B: Atomic, Molecular and Optical Physics 52~(20)
  (2019) 203001.
\newblock \href {https://doi.org/https://doi.org/10.1088/1361-6455/ab29b8}
  {\path{doi:https://doi.org/10.1088/1361-6455/ab29b8}}.
\newline\urlprefix\url{https://iopscience.iop.org/article/10.1088/1361-6455/ab29b8}

\bibitem{Misch2020}
G.~W. Misch, S.~K. Ghorui, P.~Banerjee, Y.~Sun, M.~R. Mumpower,
  \href{https://iopscience.iop.org/article/10.3847/1538-4365/abc41d}{Astromers:
  nuclear isomers in astrophysics}, The Astrophysical Journal Supplement Series
  252~(1) (2020) 2.
\newblock \href {https://doi.org/https://doi.org/10.3847/1538-4365/abc41d}
  {\path{doi:https://doi.org/10.3847/1538-4365/abc41d}}.
\newline\urlprefix\url{https://iopscience.iop.org/article/10.3847/1538-4365/abc41d}

\bibitem{Xu2004}
F.~R. Xu, E.~G. Zhao, R.~Wyss, P.~M. Walker,
  \href{https://link.aps.org/doi/10.1103/PhysRevLett.92.252501}{Enhanced
  stability of superheavy nuclei due to high-spin isomerism}, Phys. Rev. Lett.
  92 (2004) 252501.
\newblock \href {https://doi.org/10.1103/PhysRevLett.92.252501}
  {\path{doi:10.1103/PhysRevLett.92.252501}}.
\newline\urlprefix\url{https://link.aps.org/doi/10.1103/PhysRevLett.92.252501}

\bibitem{Grzywacz1998}
R.~Grzywacz, R.~B\'eraud, C.~Borcea, A.~Emsallem, M.~Glogowski, H.~Grawe,
  D.~Guillemaud-Mueller, M.~Hjorth-Jensen, M.~Houry, M.~Lewitowicz, A.~C.
  Mueller, A.~Nowak, A.~P\l{}ochocki, M.~Pf\"utzner, K.~Rykaczewski, M.~G.
  Saint-Laurent, J.~E. Sauvestre, M.~Schaefer, O.~Sorlin, J.~Szerypo,
  W.~Trinder, S.~Viteritti, J.~Winfield,
  \href{https://link.aps.org/doi/10.1103/PhysRevLett.81.766}{New island of
  $\mathit{\ensuremath{\mu}}s$ isomers in neutron-rich nuclei around the
  $\mathit{Z}\phantom{\rule{0ex}{0ex}}=\phantom{\rule{0ex}{0ex}}28$ and
  $\mathit{N}\phantom{\rule{0ex}{0ex}}=\phantom{\rule{0ex}{0ex}}40$ shell
  closures}, Phys. Rev. Lett. 81 (1998) 766--769.
\newblock \href {https://doi.org/10.1103/PhysRevLett.81.766}
  {\path{doi:10.1103/PhysRevLett.81.766}}.
\newline\urlprefix\url{https://link.aps.org/doi/10.1103/PhysRevLett.81.766}

\bibitem{nndc}
{Brookhaven National Laboratory}, National nuclear data center ({NNDC}),
  http://www.nndc.bnl.gov (2022).

\bibitem{Pedersen2023}
L.~G. Pedersen, E.~Sahin, A.~G\"orgen, F.~L. Bello~Garrote, Y.~Tsunoda,
  T.~Otsuka, M.~Niikura, S.~Nishimura, Z.~Xu, H.~Baba, G.~Benzoni, F.~Browne,
  A.~M. Bruce, S.~Ceruti, F.~C.~L. Crespi, R.~Daido, G.~de~Angelis, M.-C.
  Delattre, Z.~Dombradi, P.~Doornenbal, Y.~Fang, S.~Franchoo, G.~Gey,
  A.~Gottardo, T.~Isobe, P.~R. John, H.~S. Jung, I.~Kojouharov, T.~Kubo,
  N.~Kurz, I.~Kuti, Z.~Li, G.~Lorusso, I.~Matea, K.~Matsui, D.~Mengoni,
  T.~Miyazaki, V.~Modamio, S.~Momiyama, A.~I. Morales, P.~Morfouace, D.~R.
  Napoli, F.~Naqvi, H.~Nishibata, A.~Odahara, R.~Orlandi, Z.~Patel, S.~Rice,
  H.~Sakurai, H.~Schaffner, L.~Sinclair, P.-A. S\"oderstr\"om, D.~Sohler, I.~G.
  Stefan, T.~Sumikama, D.~Suzuki, R.~Taniuchi, J.~Taprogge, Z.~Vajta, J.~J.
  Valiente-Dob\'on, H.~Watanabe, V.~Werner, J.~Wu, A.~Yagi, M.~Yalcinkaya,
  R.~Yokoyama, K.~Yoshinaga,
  \href{https://link.aps.org/doi/10.1103/PhysRevC.107.044301}{First
  spectroscopic study of odd-odd $^{78}\mathrm{Cu}$}, Phys. Rev. C 107 (2023)
  044301.
\newblock \href {https://doi.org/10.1103/PhysRevC.107.044301}
  {\path{doi:10.1103/PhysRevC.107.044301}}.
\newline\urlprefix\url{https://link.aps.org/doi/10.1103/PhysRevC.107.044301}

\bibitem{Winger1990}
J.~A. Winger, J.~C. Hill, F.~K. Wohn, E.~K. Warburton, R.~L. Gill,
  A.~Piotrowski, R.~B. Schuhmann, D.~S. Brenner,
  \href{https://link.aps.org/doi/10.1103/PhysRevC.42.954}{Structure of
  $^{76}\mathrm{Zn}$ from $^{76}\mathrm{Cu}$ decay and systematics of
  neutron-rich zn nuclei}, Phys. Rev. C 42 (1990) 954--960.
\newblock \href {https://doi.org/10.1103/PhysRevC.42.954}
  {\path{doi:10.1103/PhysRevC.42.954}}.
\newline\urlprefix\url{https://link.aps.org/doi/10.1103/PhysRevC.42.954}

\bibitem{Kondev2021}
F.~Kondev, M.~Wang, W.~Huang, S.~Naimi, G.~Audi,
  \href{https://dx.doi.org/10.1088/1674-1137/abddae}{The nubase2020 evaluation
  of nuclear physics properties}, Chin. Phys. C 45~(3) (2021) 030001.
\newblock \href {https://doi.org/10.1088/1674-1137/abddae}
  {\path{doi:10.1088/1674-1137/abddae}}.
\newline\urlprefix\url{https://dx.doi.org/10.1088/1674-1137/abddae}

\bibitem{Canete2024}
L.~Canete, S.~Giraud, A.~Kankainen, B.~Bastin, F.~Nowacki, P.~Ascher,
  T.~Eronen, V.~{Girard Alcindor}, A.~Jokinen, A.~Khanam, I.~Moore,
  D.~Nesterenko, F.~{De Oliveira}, H.~Penttilä, C.~Petrone, I.~Pohjalainen,
  A.~{De Roubin}, V.~Rubchenya, M.~Vilen, J.~Äystö,
  \href{https://www.sciencedirect.com/science/article/pii/S0370269324002211}{Long-sought
  isomer turns out to be the ground state of 76cu}, Phys. Lett. B 853 (2024)
  138663.
\newblock \href
  {https://doi.org/https://doi.org/10.1016/j.physletb.2024.138663}
  {\path{doi:https://doi.org/10.1016/j.physletb.2024.138663}}.
\newline\urlprefix\url{https://www.sciencedirect.com/science/article/pii/S0370269324002211}

\bibitem{Roosbroeck05}
J.~{Van Roosbroeck}, H.~{De Witte}, M.~Gorska, M.~Huyse, K.~Kruglov,
  D.~Pauwels, J.-C. Thomas, K.~{Van de Vel}, P.~{Van Duppen}, S.~Franchoo,
  J.~Cederkall, V.~Fedoseyev, H.~Fynbo, U.~Georg, O.~Jonsson, U.~K{\"{o}}ster,
  L.~Weissman, W.~Mueller, V.~Mishin, D.~Fedorov, A.~{De Maesschalck},
  N.~Smirnova, K.~Heyde,
  \href{http://link.aps.org/doi/10.1103/PhysRevC.71.054307}{Evolution of the
  nuclear structure approaching $^{78}$ni: $\beta$ decay of $^{74-78}$cu},
  Phys. Rev. C 71 (2005) 054307.
\newblock \href {https://doi.org/10.1103/PhysRevC.71.054307}
  {\path{doi:10.1103/PhysRevC.71.054307}}.
\newline\urlprefix\url{http://link.aps.org/doi/10.1103/PhysRevC.71.054307}

\bibitem{Kratz1991}
K.~L. Kratz, H.~Gabelmann, P.~Möller, B.~Pfeiffer, H.~L. Ravn, A.~Wöhr, the
  ISOLDE~collaboration, Neutron-rich isotopes around the r-process
  “waiting-point” nuclei2979cu50 and3080zn50, Z. Phys. A 340 (1991).
\newblock \href {https://doi.org/https://doi.org/10.1007/BF01290331}
  {\path{doi:https://doi.org/10.1007/BF01290331}}.

\bibitem{Guenaut2007}
C.~Gu\'enaut, G.~Audi, D.~Beck, K.~Blaum, G.~Bollen, P.~Delahaye, F.~Herfurth,
  A.~Kellerbauer, H.-J. Kluge, J.~Libert, D.~Lunney, S.~Schwarz,
  L.~Schweikhard, C.~Yazidjian,
  \href{https://link.aps.org/doi/10.1103/PhysRevC.75.044303}{High-precision
  mass measurements of nickel, copper, and gallium isotopes and the purported
  shell closure at $n=40$}, Phys. Rev. C 75 (2007) 044303.
\newblock \href {https://doi.org/10.1103/PhysRevC.75.044303}
  {\path{doi:10.1103/PhysRevC.75.044303}}.
\newline\urlprefix\url{https://link.aps.org/doi/10.1103/PhysRevC.75.044303}

\bibitem{Winger2009}
J.~A. Winger, S.~V. Ilyushkin, K.~P. Rykaczewski, C.~J. Gross, J.~C.
  Batchelder, C.~Goodin, R.~Grzywacz, J.~H. Hamilton, A.~Korgul, W.~Kr\'olas,
  S.~N. Liddick, C.~Mazzocchi, S.~Padgett, A.~Piechaczek, M.~M. Rajabali,
  D.~Shapira, E.~F. Zganjar, I.~N. Borzov,
  \href{https://link.aps.org/doi/10.1103/PhysRevLett.102.142502}{Large
  $\ensuremath{\beta}$-delayed neutron emission probabilities in the
  $^{78}\mathrm{Ni}$ region}, Phys. Rev. Lett. 102 (2009) 142502.
\newblock \href {https://doi.org/10.1103/PhysRevLett.102.142502}
  {\path{doi:10.1103/PhysRevLett.102.142502}}.
\newline\urlprefix\url{https://link.aps.org/doi/10.1103/PhysRevLett.102.142502}

\bibitem{Hosmer2010}
P.~Hosmer, H.~Schatz, A.~Aprahamian, O.~Arndt, R.~R.~C. Clement, A.~Estrade,
  K.~Farouqi, K.-L. Kratz, S.~N. Liddick, A.~F. Lisetskiy, P.~F. Mantica,
  P.~M\"oller, W.~F. Mueller, F.~Montes, A.~C. Morton, M.~Ouellette,
  E.~Pellegrini, J.~Pereira, B.~Pfeiffer, P.~Reeder, P.~Santi, M.~Steiner,
  A.~Stolz, B.~E. Tomlin, W.~B. Walters, A.~W\"ohr,
  \href{https://link.aps.org/doi/10.1103/PhysRevC.82.025806}{Half-lives and
  branchings for $\ensuremath{\beta}$-delayed neutron emission for neutron-rich
  co--cu isotopes in the $r$-process}, Phys. Rev. C 82 (2010) 025806.
\newblock \href {https://doi.org/10.1103/PhysRevC.82.025806}
  {\path{doi:10.1103/PhysRevC.82.025806}}.
\newline\urlprefix\url{https://link.aps.org/doi/10.1103/PhysRevC.82.025806}

\bibitem{Welker2017}
A.~Welker, N.~A.~S. Althubiti, D.~Atanasov, K.~Blaum, T.~E. Cocolios,
  F.~Herfurth, S.~Kreim, D.~Lunney, V.~Manea, M.~Mougeot, D.~Neidherr,
  F.~Nowacki, A.~Poves, M.~Rosenbusch, L.~Schweikhard, F.~Wienholtz, R.~N.
  Wolf, K.~Zuber,
  \href{https://link.aps.org/doi/10.1103/PhysRevLett.119.192502}{Binding energy
  of $^{79}\mathrm{Cu}$: Probing the structure of the doubly magic
  $^{78}\mathrm{Ni}$ from only one proton away}, Phys. Rev. Lett. 119 (2017)
  192502.
\newblock \href {https://doi.org/10.1103/PhysRevLett.119.192502}
  {\path{doi:10.1103/PhysRevLett.119.192502}}.
\newline\urlprefix\url{https://link.aps.org/doi/10.1103/PhysRevLett.119.192502}

\bibitem{Groote2017}
R.~P. de~Groote, J.~Billowes, C.~L. Binnersley, M.~L. Bissell, T.~E. Cocolios,
  T.~Day~Goodacre, G.~J. Farooq-Smith, D.~V. Fedorov, K.~T. Flanagan,
  S.~Franchoo, R.~F. Garcia~Ruiz, A.~Koszor\'us, K.~M. Lynch, G.~Neyens,
  F.~Nowacki, T.~Otsuka, S.~Rothe, H.~H. Stroke, Y.~Tsunoda, A.~R. Vernon,
  K.~D.~A. Wendt, S.~G. Wilkins, Z.~Y. Xu, X.~F. Yang,
  \href{https://link.aps.org/doi/10.1103/PhysRevC.96.041302}{Dipole and
  quadrupole moments of $^{73\text{--}78}\mathrm{Cu}$ as a test of the
  robustness of the $z=28$ shell closure near $^{78}\mathrm{Ni}$}, Phys. Rev. C
  96 (2017) 041302.
\newblock \href {https://doi.org/10.1103/PhysRevC.96.041302}
  {\path{doi:10.1103/PhysRevC.96.041302}}.
\newline\urlprefix\url{https://link.aps.org/doi/10.1103/PhysRevC.96.041302}

\bibitem{Silwal2022}
U.~Silwal, J.~A. Winger, S.~V. Ilyushkin, K.~P. Rykaczewski, C.~J. Gross, J.~C.
  Batchelder, L.~Cartegni, I.~G. Darby, R.~Grzywacz, A.~Korgul, W.~Kr\'olas,
  S.~N. Liddick, C.~Mazzocchi, A.~J. Mendez, S.~Padgett, M.~M. Rajabali, D.~P.
  Siwakoti, D.~Shapira, D.~W. Stracener, E.~F. Zganjar,
  \href{https://link.aps.org/doi/10.1103/PhysRevC.106.044311}{$\ensuremath{\beta}$
  decay of neutron-rich $^{76}\mathrm{Cu}$ and the structure of
  $^{76}\mathrm{Zn}$}, Phys. Rev. C 106 (2022) 044311.
\newblock \href {https://doi.org/10.1103/PhysRevC.106.044311}
  {\path{doi:10.1103/PhysRevC.106.044311}}.
\newline\urlprefix\url{https://link.aps.org/doi/10.1103/PhysRevC.106.044311}

\bibitem{Pedersen2024}
L.~G. {Pedersen}, \href{https://hdl.handle.net/10852/108681}{{Nuclear structure
  along the Z = 28 and Z = 50 shells - Shell evolution and shape coexistence in
  $^{74,76,78}$Cu and $^{126}$Sn}}, Ph.D. thesis, University of Oslo (2024).
\newline\urlprefix\url{https://hdl.handle.net/10852/108681}

\bibitem{Giraud2022}
S.~Giraud, L.~Canete, B.~Bastin, A.~Kankainen, A.~Fantina, F.~Gulminelli,
  P.~Ascher, T.~Eronen, V.~Girard-Alcindor, A.~Jokinen, A.~Khanam, I.~Moore,
  D.~Nesterenko, F.~{de Oliveira Santos}, H.~Penttilä, C.~Petrone,
  I.~Pohjalainen, A.~{De Roubin}, V.~Rubchenya, M.~Vilen, J.~Äystö,
  \href{https://www.sciencedirect.com/science/article/pii/S0370269322004439}{Mass
  measurements towards doubly magic 78ni: Hydrodynamics versus nuclear mass
  contribution in core-collapse supernovae}, Physics Letters B 833 (2022)
  137309.
\newblock \href
  {https://doi.org/https://doi.org/10.1016/j.physletb.2022.137309}
  {\path{doi:https://doi.org/10.1016/j.physletb.2022.137309}}.
\newline\urlprefix\url{https://www.sciencedirect.com/science/article/pii/S0370269322004439}

\bibitem{Singh2024}
B.~Singh, J.~Chen, A.~R. Farhan,
  \href{https://www.sciencedirect.com/science/article/pii/S0090375224000176}{Nuclear
  structure and decay data for a=76 isobars}, Nucl. Data Sheets 194 (2024)
  3--459.
\newblock \href {https://doi.org/https://doi.org/10.1016/j.nds.2024.02.002}
  {\path{doi:https://doi.org/10.1016/j.nds.2024.02.002}}.
\newline\urlprefix\url{https://www.sciencedirect.com/science/article/pii/S0090375224000176}

\bibitem{Tolosa2019}
A.~Tolosa-Delgado, J.~Agramunt, J.~Tain, A.~Algora, C.~Domingo-Pardo,
  A.~Morales, B.~Rubio, A.~Tarifeño-Saldivia, F.~Calviño, G.~Cortes,
  N.~Brewer, B.~Rasco, K.~Rykaczewski, D.~Stracener, J.~Allmond, R.~Grzywacz,
  R.~Yokoyama, M.~Singh, T.~King, M.~Madurga, S.~Nishimura, V.~Phong, S.~Go,
  J.~Liu, K.~Matsui, H.~Sakurai, G.~Kiss, T.~Isobe, H.~Baba, S.~Kubono,
  N.~Fukuda, D.~Ahn, Y.~Shimizu, T.~Sumikama, H.~Suzuki, H.~Takeda,
  P.~Söderström, M.~Takechi, C.~Bruno, T.~Davinson, C.~Griffin, O.~Hall,
  D.~Kahl, P.~Woods, P.~Coleman-Smith, M.~Labiche, I.~Lazarus, P.~Morrall,
  V.~E. Pucknell, J.~Simpson, S.~Thomas, M.~Prydderch, L.~Harkness-Brennan,
  R.~Page, I.~Dillmann, R.~Caballero-Folch, Y.~Saito, A.~Estrade, N.~Nepal,
  F.~Montes, G.~Lorusso, J.~Liang, S.~Bae, J.~Ha, B.~Moon,
  \href{https://www.sciencedirect.com/science/article/pii/S0168900219301743}{Commissioning
  of the briken detector for the measurement of very exotic $\beta$-delayed
  neutron emitters}, Nucl. Instrum. Methods Phys. Res. A 925 (2019) 133--147.
\newblock \href {https://doi.org/https://doi.org/10.1016/j.nima.2019.02.004}
  {\path{doi:https://doi.org/10.1016/j.nima.2019.02.004}}.
\newline\urlprefix\url{https://www.sciencedirect.com/science/article/pii/S0168900219301743}

\bibitem{Tolosa2020}
A.~{Tolosa-Delgado},
  \href{https://roderic.uv.es/items/599dc1ca-3aa7-43ce-806d-681ce15c9d36}{{Study
  of $\beta$-delayed neutron emitters in the region of $^{78}$Ni and its impact
  on r-process nucleosynthesis}}, Ph.D. thesis, IFIC-CSIC (2020).
\newline\urlprefix\url{https://roderic.uv.es/items/599dc1ca-3aa7-43ce-806d-681ce15c9d36}

\bibitem{Nutaq}
N.~data~acquisition system, https://www.nutaq.com/ (2022).

\bibitem{Benito2020}
J.~Benito, L.~M. Fraile, A.~Korgul, M.~Piersa, E.~Adamska, A.~N. Andreyev,
  R.~\'Alvarez-Rodr\'{\i}guez, A.~E. Barzakh, G.~Benzoni, T.~Berry, M.~J.~G.
  Borge, M.~Carmona, K.~Chrysalidis, C.~Costache, J.~G. Cubiss,
  T.~Day~Goodacre, H.~De~Witte, D.~V. Fedorov, V.~N. Fedosseev,
  G.~Fern\'andez-Mart\'{\i}nez, A.~Fija\l{}kowska, M.~Fila, H.~Fynbo,
  D.~Galaviz, P.~Galve, M.~Garc\'{\i}a-D\'{\i}ez, P.~T. Greenlees, R.~Grzywacz,
  L.~J. Harkness-Brennan, C.~Henrich, M.~Huyse, P.~Ib\'a\~nez, A.~Illana,
  Z.~Janas, J.~Jolie, D.~S. Judson, V.~Karayonchev, M.~Kici\ifmmode
  \acute{n}\else~\'{n}\fi{}ska Habior, J.~Konki, J.~Kurcewicz, I.~Lazarus,
  R.~Lic\ifmmode~\u{a}\else \u{a}\fi{}, A.~L\'opez-Montes, M.~Lund, H.~Mach,
  M.~Madurga, I.~Marroqu\'{\i}n, B.~Marsh, M.~C. Mart\'{\i}nez, C.~Mazzocchi,
  N.~M\ifmmode~\u{a}\else \u{a}\fi{}rginean, R.~M\ifmmode~\u{a}\else
  \u{a}\fi{}rginean, K.~Miernik, C.~Mihai, R.~E. Mihai, E.~N\'acher, A.~Negret,
  B.~Olaizola, R.~D. Page, S.~V. Paulauskas, S.~Pascu, A.~Perea, V.~Pucknell,
  P.~Rahkila, C.~Raison, E.~Rapisarda, J.-M. R\'egis, K.~Rezynkina, F.~Rotaru,
  S.~Rothe, D.~S\'anchez-Parcerisa, V.~S\'anchez-Tembleque, K.~Schomacker,
  G.~S. Simpson, C.~Sotty, L.~Stan, M.~St\ifmmode~\u{a}\else \u{a}\fi{}noiu,
  M.~Stryjczyk, O.~Tengblad, A.~Turturica, J.~M. Ud\'{\i}as, P.~Van~Duppen,
  V.~Vedia, A.~Villa-Abaunza, S.~Vi\~nals, W.~B. Walters, R.~Wadsworth,
  N.~Warr, \href{https://link.aps.org/doi/10.1103/PhysRevC.102.014328}{Detailed
  spectroscopy of doubly magic $^{132}\mathrm{Sn}$}, Phys. Rev. C 102 (2020)
  014328.
\newblock \href {https://doi.org/10.1103/PhysRevC.102.014328}
  {\path{doi:10.1103/PhysRevC.102.014328}}.
\newline\urlprefix\url{https://link.aps.org/doi/10.1103/PhysRevC.102.014328}

\bibitem{Piersa2021}
M.~Piersa-Si\l{}kowska, A.~Korgul, J.~Benito, L.~M. Fraile, E.~Adamska, A.~N.
  Andreyev, R.~\'Alvarez-Rodr\'{\i}guez, A.~E. Barzakh, G.~Benzoni, T.~Berry,
  M.~J.~G. Borge, M.~Carmona, K.~Chrysalidis, J.~G. Correia, C.~Costache, J.~G.
  Cubiss, T.~Day~Goodacre, H.~De~Witte, D.~V. Fedorov, V.~N. Fedosseev,
  G.~Fern\'andez-Mart\'{\i}nez, A.~Fija\l{}kowska, H.~Fynbo, D.~Galaviz,
  P.~Galve, M.~Garc\'{\i}a-D\'{\i}ez, P.~T. Greenlees, R.~Grzywacz, L.~J.
  Harkness-Brennan, C.~Henrich, M.~Huyse, P.~Ib\'a\~nez, A.~Illana, Z.~Janas,
  K.~Johnston, J.~Jolie, D.~S. Judson, V.~Karanyonchev, M.~Kici\ifmmode
  \acute{n}\else~\'{n}\fi{}ska Habior, J.~Konki, L.~Koszuk, J.~Kurcewicz,
  I.~Lazarus, R.~Lic\ifmmode~\u{a}\else \u{a}\fi{}, A.~L\'opez-Montes, H.~Mach,
  M.~Madurga, I.~Marroqu\'{\i}n, B.~Marsh, M.~C. Mart\'{\i}nez, C.~Mazzocchi,
  K.~Miernik, C.~Mihai, N.~M\ifmmode~\u{a}\else \u{a}\fi{}rginean,
  R.~M\ifmmode~\u{a}\else \u{a}\fi{}rginean, A.~Negret, E.~N\'acher, J.~Ojala,
  B.~Olaizola, R.~D. Page, J.~Pakarinen, S.~Pascu, S.~V. Paulauskas, A.~Perea,
  V.~Pucknell, P.~Rahkila, C.~Raison, E.~Rapisarda, K.~Rezynkina, F.~Rotaru,
  S.~Rothe, K.~P. Rykaczewski, J.-M. R\'egis, K.~Schomacker, M.~Si\l{}kowski,
  G.~Simpson, C.~Sotty, L.~Stan, M.~St\ifmmode~\u{a}\else \u{a}\fi{}noiu,
  M.~Stryjczyk, D.~S\'anchez-Parcerisa, V.~S\'anchez-Tembleque, O.~Tengblad,
  A.~Turturic\ifmmode~\u{a}\else \u{a}\fi{}, J.~M. Ud\'{\i}as, P.~Van~Duppen,
  V.~Vedia, A.~Villa, S.~Vi\~nals, R.~Wadsworth, W.~B. Walters, N.~Warr, S.~G.
  Wilkins, \href{https://link.aps.org/doi/10.1103/PhysRevC.104.044328}{First
  $\ensuremath{\beta}$-decay spectroscopy of $^{135}\mathrm{In}$ and new
  $\ensuremath{\beta}$-decay branches of $^{134}\mathrm{In}$}, Phys. Rev. C 104
  (2021) 044328.
\newblock \href {https://doi.org/10.1103/PhysRevC.104.044328}
  {\path{doi:10.1103/PhysRevC.104.044328}}.
\newline\urlprefix\url{https://link.aps.org/doi/10.1103/PhysRevC.104.044328}

\bibitem{GRAPENGIESSER1974}
B.~Grapengiesser, E.~Lund, G.~Rudstam,
  \href{https://www.sciencedirect.com/science/article/pii/0022190274804483}{Survey
  of short-lived fission products obtained using the isotope-separator-on-line
  facility at studsvik}, J. Inorg. and Nucl. Chem. 36~(11) (1974) 2409--2431.
\newblock \href {https://doi.org/https://doi.org/10.1016/0022-1902(74)80448-3}
  {\path{doi:https://doi.org/10.1016/0022-1902(74)80448-3}}.
\newline\urlprefix\url{https://www.sciencedirect.com/science/article/pii/0022190274804483}

\bibitem{Koester2000}
U.~Koester, \href{https://cds.cern.ch/record/494272}{Yields and spectroscopy of
  radioactive isotopes at lohengrin and isolde; ausbeuten und spektroskopie
  radioaktiver isotope bei lohengrin und isolde}, Ph.D. thesis, echnische
  Universität München (Jul 2000).
\newline\urlprefix\url{https://cds.cern.ch/record/494272}

\bibitem{Hukkanen2023}
M.~Hukkanen, W.~Ryssens, P.~Ascher, M.~Bender, T.~Eronen, S.~Gr\'evy,
  A.~Kankainen, M.~Stryjczyk, L.~Al~Ayoubi, S.~Ayet, O.~Beliuskina,
  C.~Delafosse, W.~Gins, M.~Gerbaux, A.~Husson, A.~Jokinen, D.~A. Nesterenko,
  I.~Pohjalainen, M.~Reponen, S.~Rinta-Antila, A.~de~Roubin, A.~P. Weaver,
  \href{https://link.aps.org/doi/10.1103/PhysRevC.107.014306}{Odd-odd
  neutron-rich rhodium isotopes studied with the double penning trap jyfltrap},
  Phys. Rev. C 107 (2023) 014306.
\newblock \href {https://doi.org/10.1103/PhysRevC.107.014306}
  {\path{doi:10.1103/PhysRevC.107.014306}}.
\newline\urlprefix\url{https://link.aps.org/doi/10.1103/PhysRevC.107.014306}

\bibitem{GrootePC}
R.~P. de~Groote, Private communication. (2024).

\bibitem{Turkat2023}
S.~Turkat, X.~Mougeot, B.~Singh, K.~Zuber,
  \href{https://www.sciencedirect.com/science/article/pii/S0092640X23000128}{Systematics
  of logft values for $\beta^-$, and ec/$\beta^+$transitions}, At. Data Nucl.
  Data Tables 152 (2023) 101584.
\newblock \href {https://doi.org/https://doi.org/10.1016/j.adt.2023.101584}
  {\path{doi:https://doi.org/10.1016/j.adt.2023.101584}}.
\newline\urlprefix\url{https://www.sciencedirect.com/science/article/pii/S0092640X23000128}

\bibitem{Hardy1977}
J.~Hardy, L.~Carraz, B.~Jonson, P.~Hansen,
  \href{https://www.sciencedirect.com/science/article/pii/0370269377902234}{The
  essential decay of pandemonium: A demonstration of errors in complex
  beta-decay schemes}, Phys. Lett. B 71~(2) (1977) 307--310.
\newblock \href {https://doi.org/https://doi.org/10.1016/0370-2693(77)90223-4}
  {\path{doi:https://doi.org/10.1016/0370-2693(77)90223-4}}.
\newline\urlprefix\url{https://www.sciencedirect.com/science/article/pii/0370269377902234}

\bibitem{WaltersPC}
W.~B. Walters, Private communication. (2024).

\bibitem{Hagberg1997}
E.~Hagberg, I.~Towner, J.~Hardy, V.~Koslowsky, G.~Savard, S.~Sterbenz,
  \href{https://www.sciencedirect.com/science/article/pii/S0375947496004320}{Beta
  decays of 44v and 52co}, Nucl. Phys. A 613~(3) (1997) 183--198.
\newblock \href {https://doi.org/https://doi.org/10.1016/S0375-9474(96)00432-0}
  {\path{doi:https://doi.org/10.1016/S0375-9474(96)00432-0}}.
\newline\urlprefix\url{https://www.sciencedirect.com/science/article/pii/S0375947496004320}

\bibitem{Pauwels2009}
D.~Pauwels, O.~Ivanov, N.~Bree, J.~B\"uscher, T.~E. Cocolios, M.~Huyse,
  Y.~Kudryavtsev, R.~Raabe, M.~Sawicka, J.~V. de~Walle, P.~V. Duppen,
  A.~Korgul, I.~Stefanescu, A.~A. Hecht, N.~Hoteling, A.~W\"ohr, W.~B. Walters,
  R.~Broda, B.~Fornal, W.~Krolas, T.~Pawlat, J.~Wrzesinski, M.~P. Carpenter,
  R.~V.~F. Janssens, T.~Lauritsen, D.~Seweryniak, S.~Zhu, J.~R. Stone, X.~Wang,
  \href{https://link.aps.org/doi/10.1103/PhysRevC.79.044309}{Structure of
  $^{65,67}\mathrm{Co}$ studied through the $\ensuremath{\beta}$ decay of
  $^{65,67}\mathrm{Fe}$ and a deep-inelastic reaction}, Phys. Rev. C 79 (2009)
  044309.
\newblock \href {https://doi.org/10.1103/PhysRevC.79.044309}
  {\path{doi:10.1103/PhysRevC.79.044309}}.
\newline\urlprefix\url{https://link.aps.org/doi/10.1103/PhysRevC.79.044309}

\bibitem{Cerny1972}
J.~Cerny, R.~Gough, R.~Sextro, J.~E. Esterl,
  \href{https://www.sciencedirect.com/science/article/pii/0375947472902266}{Further
  results on the proton radioactivity of 53mco}, Nucl. Phys. A 188~(3) (1972)
  666--672.
\newblock \href {https://doi.org/https://doi.org/10.1016/0375-9474(72)90226-6}
  {\path{doi:https://doi.org/10.1016/0375-9474(72)90226-6}}.
\newline\urlprefix\url{https://www.sciencedirect.com/science/article/pii/0375947472902266}

\bibitem{Roosbroeck2004}
J.~Van~Roosbroeck, H.~De~Witte, M.~Gorska, M.~Huyse, K.~Kruglov, K.~Van~de Vel,
  P.~Van~Duppen, S.~Franchoo, J.~Cederkall, V.~N. Fedoseyev, H.~Fynbo,
  U.~Georg, O.~Jonsson, U.~K\"oster, L.~Weissman, W.~F. Mueller, V.~I. Mishin,
  D.~Fedorov, W.~B. Walters, N.~A. Smirnova, A.~Van~Dyck, A.~De~Maesschalck,
  K.~Heyde, \href{https://link.aps.org/doi/10.1103/PhysRevC.69.034313}{Coupling
  a proton and a neutron to the semidoubly magic nucleus $^{68}\mathrm{Ni}$: A
  study of $^{70}\mathrm{Cu}$ via the $\ensuremath{\beta}$ decay of
  $^{70}\mathrm{Ni}$ and $^{70}\mathrm{Cu}$}, Phys. Rev. C 69 (2004) 034313.
\newblock \href {https://doi.org/10.1103/PhysRevC.69.034313}
  {\path{doi:10.1103/PhysRevC.69.034313}}.
\newline\urlprefix\url{https://link.aps.org/doi/10.1103/PhysRevC.69.034313}

\bibitem{Paziy2020}
V.~Paziy, L.~M. Fraile, H.~Mach, B.~Olaizola, G.~S. Simpson, A.~Aprahamian,
  C.~Bernards, J.~A. Briz, B.~Bucher, C.~J. Chiara, Z.~Dlouh\'y, I.~Gheorghe,
  D.~Ghi\ifmmode \mbox{\c{t}}\else \c{t}\fi{}\ifmmode~\check{a}\else
  \v{a}\fi{}, P.~Hoff, J.~Jolie, U.~K\"oster, W.~Kurcewicz,
  R.~Lic\ifmmode~\check{a}\else \v{a}\fi{}, N.~M\ifmmode~\check{a}\else
  \v{a}\fi{}rginean, R.~M\ifmmode~\check{a}\else \v{a}\fi{}rginean, J.-M.
  R\'egis, M.~Rudigier, T.~Sava, M.~St\ifmmode~\check{a}\else \v{a}\fi{}noiu,
  L.~Stroe, W.~B. Walters,
  \href{https://link.aps.org/doi/10.1103/PhysRevC.102.014329}{Fast-timing study
  of $^{81}\mathrm{Ga}$ from the $\ensuremath{\beta}$ decay of
  $^{81}\mathrm{Zn}$}, Phys. Rev. C 102 (2020) 014329.
\newblock \href {https://doi.org/10.1103/PhysRevC.102.014329}
  {\path{doi:10.1103/PhysRevC.102.014329}}.
\newline\urlprefix\url{https://link.aps.org/doi/10.1103/PhysRevC.102.014329}

\bibitem{Kibedi2008}
T.~Kibédi, T.~Burrows, M.~Trzhaskovskaya, P.~Davidson, C.~Nestor,
  \href{https://www.sciencedirect.com/science/article/pii/S0168900208002520}{Evaluation
  of theoretical conversion coefficients using bricc}, Nucl. Instrum. Methods
  Phys. Res. A 589~(2) (2008) 202--229.
\newblock \href {https://doi.org/https://doi.org/10.1016/j.nima.2008.02.051}
  {\path{doi:https://doi.org/10.1016/j.nima.2008.02.051}}.
\newline\urlprefix\url{https://www.sciencedirect.com/science/article/pii/S0168900208002520}

\bibitem{Chester21}
A.~Chester, B.~A. Brown, S.~P. Burcher, M.~P. Carpenter, J.~J. Carroll, C.~J.
  Chiara, P.~A. Copp, B.~P. Crider, J.~T. Harke, D.~E.~M. Hoff, K.~Kolos, S.~N.
  Liddick, B.~Longfellow, M.~J. Mogannam, T.~H. Ogunbeku, C.~J. Prokop,
  D.~Rhodes, A.~L. Richard, O.~A. Shehu, A.~S. Tamashiro, R.~Unz, Y.~Xiao,
  \href{https://link.aps.org/doi/10.1103/PhysRevC.104.054314}{Identification of
  a new isomeric state in $^{76}\mathrm{Zn}$ following the $\ensuremath{\beta}$
  decay of $^{76}\mathrm{Cu}$}, Phys. Rev. C 104 (2021) 054314.
\newblock \href {https://doi.org/10.1103/PhysRevC.104.054314}
  {\path{doi:10.1103/PhysRevC.104.054314}}.
\newline\urlprefix\url{https://link.aps.org/doi/10.1103/PhysRevC.104.054314}

\bibitem{Pedersen2019}
L.~G. {Pedersen},
  \href{https://www.duo.uio.no/bitstream/handle/10852/69267/Line_Pedersen_masteroppgave.pdf}{{Shell
  evolution towards $^{78}$Ni: Spectroscopy of $^{74}$Cu and $^{76}$Cu}},
  Master's thesis, University of Oslo (2019).
\newline\urlprefix\url{https://www.duo.uio.no/bitstream/handle/10852/69267/Line_Pedersen_masteroppgave.pdf}

\end{thebibliography}


\end{document}